\title{\huge A Bayesian decision-theoretic approach to incorporate preclinical information into phase I oncology trials}
\author{
        Haiyan Zheng \\
		 Institute of Health and Society, Newcastle University, U.K. \\
		 \href{mailto:haiyan.zheng@newcastle.ac.uk}{haiyan.zheng@newcastle.ac.uk} \\
		 \vspace{1em} \\
        Lisa V. Hampson \\
        Statistical Methodology and Consulting, Novartis Pharma AG, Switzerland.
}
\date{ }

\documentclass[11pt]{article}

\usepackage[margin=0.78in]{geometry}
\usepackage{hyperref}
\usepackage{amssymb, amsmath}
\usepackage{graphicx, caption}
\usepackage{booktabs}

\usepackage{natbib}

\usepackage{color,tikz}
\usetikzlibrary{shapes,arrows}

\usepackage{pdfpages}

\begin{document}
\maketitle

\begin{abstract}
Leveraging preclinical animal data for a phase I first-in-man trial is appealing yet challenging. A prior based on animal data may place large probability mass on values of the dose-toxicity model parameter(s), which appear infeasible in light of data accrued from the ongoing phase I clinical trial. In this paper, we seek to use animal data to improve decision making in a model-based dose-escalation procedure for phase I oncology trials. Specifically, animal data are incorporated via a robust mixture prior for the parameters of the dose-toxicity relationship. This prior changes dynamically as the trial progresses. After completion of treatment for each cohort, the weight allocated to the informative component, obtained based on animal data alone, is updated using a decision-theoretic approach to assess the commensurability of the animal data with the human toxicity data observed thus far. In particular, we measure commensurability as a function of the utility of optimal prior predictions for the human responses (toxicity or no toxicity) on each administered dose. The proposed methodology is illustrated through several examples and an extensive simulation study. Results show that our proposal can address difficulties in coping with prior-data conflict commencing in sequential trials with a small sample size.
\end{abstract}

\section{Introduction}

Phase I oncology trials are performed primarily to characterise the toxicity profile of an anticancer therapy in human subjects. Regulatory authorities require that first-in-man trials be preceded by non-clinical safety studies in at least two animal species, including the most sensitive and relevant species  \citep{FDA:FIH2005, EMA:FIH2008}.
It is therefore reasonable to assume that some animal data will be available at the time of designing a phase I clinical trial. Despite this availability, trialists may be uncertain about how and whether to leverage animal data. On the one hand, there is considerable motivation to design and conduct more efficient phase I clinical trials, basing interim dose-escalation decisions on relevant information internal and external to the trial \citep{BIOM:JO1990, SBR:Neuenschwander2016}. 
On the other hand, leveraging animal data could jeopardise patient safety \citep{JRSSA:Senn2007, JLME:Dresser2009, Lancet:Balkwill2011}
if parameters of the human dose-toxicity relationship, predicted from animal data, are non-exchangeable with those of the true (but unknown) human relationship. 
In this paper, we propose a quantitative Bayesian approach to leveraging preclinical animal data, for the design and interpretation of a phase I first-in-man oncology trial. \\

There exist Bayesian and frequentist approaches for incorporating historical data into the analysis of a new clinical trial \citep{TIRS:Lim2018}. This paper will focus on Bayesian approaches for leveraging historical, or more generally, complementary (co-)data  \citep{SBR:Neuenschwander2016}, comprising all relevant historical and concurrent data into a new study. 
Power priors offer an approach to using external information: they raise the likelihood of the co-data to a power between 0 and 1. In particular, this exponent can be treated as a constant and could be fixed in advance on the basis of our prior scientific understanding of the similarity of the parameters underpinning the distributions of the co- and new data. \citet{PS:Gravestock2017} propose adaptive versions of the power prior, which fix the power equal to the estimate maximising the marginal likelihood of the historical and new data.
Alternatively, the normalised power priors \citep{JABES:Duan2006} treat the power weight as a random variable, with a prior distribution which is updated once the new data are observed. 
The degree of down-weighting reacts to the similarity of the co-data and new trial data; for this reason, they are referred to as Bayesian dynamic priors. \\

Commensurate priors \citep{BIOM:Hobbs2011, BA:Hobbs2012} and meta-analytic predictive priors \citep{CT:Neuenschwander2010, BIOM:Heinz2014} are other examples of Bayesian approaches to dynamic borrowing. Meta-analytic approaches assume that study-specific parameters are exchangeable, that is,
parameters are thought of as conditionally independent draws from, for example, a normal distribution with unknown mean $\mu$ and variance $\tau^2$ \citep{JRSSA:Higgins2009}. Here $\tau$ captures the degree of between-study heterogeneity, with larger values leading to greater attenuation of the co-data. 
To encourage faster discounting of the co-data in cases of a prior-data conflict, \citet{BIOM:Heinz2014} develop heavy-tailed prior \citep{BA:O'Hagan2012}, which is a two-component mixture distribution with one component the meta-analytic predictive prior and the other a weakly-informative (proper) distribution.
\citet{PS:Mutsvari2016} discuss how to choose the weakly-informative component to ensure appropriate down-weighting behaviour when there is a conflict.  \\

These Bayesian approaches have mainly found applications in phase II clinical trials to use historical controls to increase power and precision \citep{PS:Viele2014}. 
Comparatively little has been written on leveraging historical data, particularly preclinical animal data, for phase I first-in-man trials, where different metrics are needed to evaluate the benefits of borrowing on trial operating characteristics. 
\citet{SiM:Zheng2018} propose a robust Bayesian hierarchical random-effects model to incorporate animal data from one or multiple species into a phase I oncology trial. 
Informed by allometric scaling, translation parameter is used to bring animal data onto an equivalent human dosing scale. 
A log-normal prior is specified for the translation parameter to quantify our uncertainty about the magnitude of differences between dose-toxicity model parameters in animals and humans.
In light of this scaling, it is reasonable to assume parameters of the translated animal dose-toxicity model parameters are exchangeable with parameters in humans. \\

In this paper, we represent animal data in a prior for parameters of the human dose-toxicity relationship, assuming this can be described by a two-parameter logistic regression model \citep{BioPS:Whitehead1998, SIM:Neuenschwander2008}.  We analyse the accumulating human data on dose-limiting toxicities (DLTs) using a robust two-component mixture prior. We use empirical prior mixture weights which are dynamically updated as the trial progresses to reflect the degree of agreement between the observed human data and predictions based on animal data. We adopt a Bayesian decision-theoretic approach, defining the degree of agreement as the attained utility (termed here as predictive utility) of the animal predictions, averaging across the relevant doses used thus far in the trial. After completion of each cohort, the mixture prior with empirically chosen weights is updated with the observed human data to derive a posterior distribution for dose-toxicity model parameters, which is used to derive dose recommendations for the next patient cohort. Bayesian decision theory has been previously used for the design of early phase clinical trials \citep{SiM:Whitehead1995, BioPS:Whitehead1998, BIOS:Whitehead2001, SiM:Stallard2001}. \citet{SMMR:Hee2016} review Bayesian decision-theoretic approaches for designing small trials and pilot studies. \\

The remainder of the paper is structured as follows. Starting with a motivating example in Section \ref{sec:motivate}, we explain in Section \ref{sec:BayesMod} how animal data from a single species on two or more doses can be represented in a bivariate normal prior for the dose-toxicity parameters that underpin the human trial. 
In Section \ref{sec:mixprior}, we describe the Bayesian decision-theoretic approach used to measure the degree of agreement between human DLT outcomes and what we would expect based on the animal data. We retrospectively design and analyse the example trial applying the proposed methodology in section \ref{sec:examplePhI}, and describe a simulation study performed to evaluate the operating characteristics of a Bayesian phase I dose-escalation study in Section \ref{sec:sims}. 
We close with a discussion of our findings and future research in Section \ref{sec:discuss}.

\section{Motivating example}
\label{sec:motivate}

\citet{AACR:Sessa2013} report a phase I first-in-man trial intended to estimate the maximum tolerated dose (MTD) of the anticancer therapy AUY922. The trial enrolled 101 patients who were treated sequentially in cohorts of size three or more. Patients were monitored to record whether or not they experienced a dose-limiting toxicity (DLT). Nine doses were evaluated during the study, summarised by the dosing set $\mathcal{D}=\{ 2, 4, 8, 16, 22, 28, 40, 54, 70\}$ mg/m$^2$. 
Dose-escalation decisions were guided by a two-parameter Bayesian logistic regression model (BLRM) for the relationship between dose and DLT risk \citep{SIM:Neuenschwander2008}.
The model was implemented placing weakly informative priors on the model parameters consistent with median DLT probabilities of around 0.1\% and 33\% at doses 2 and 28 mg/m$^2$, respectively; these settings were informed by data from toxicology studies performed in dogs \citep{AACR:Sessa2013}, although the animal data were not formally incorporated into the analysis of the phase I trial.
Let $p_i$ denote the DLT risk on the $i$th largest dose in $\mathcal{D}$, $d_i$, and let $\boldsymbol{x}_\mathcal{H}^{(h-1)}$ be a vector of observed outcomes from the first $(h-1)$ cohorts of patients. 
\citet{AACR:Sessa2013} implemented the BLRM-guided escalation procedure to recommend that cohort $h$ receives dose $d_\text{sel}^{(h)}$, defined as the highest dose in $\mathcal{D}$ with an acceptable risk of excessive toxicity:
\begin{equation}
\label{eq:overdose}
d_\text{sel}^{(h)} = \max \{d_i \in \mathcal{D}:  \text{Pr}(p_i \geq 0.33 | \boldsymbol{x}_\mathcal{H}^{(h-1)}) \leq 0.25\}.
\end{equation}
\noindent \citet{AACR:Sessa2013} implemented the BLRM-guided escalation procedure with an additional constraint that $d_\text{sel}^{(h)}$ should not exceed more than a two-fold increase in the current dose. \\

The AUY922 phase I clinical trial prompts the following questions:
\begin{itemize}
\item[(i)] How can we formally incorporate preclinical data into prior distributions for the dose-toxicity model parameters?
\item[(ii)] How can we dynamically update these priors in response to any prior-data conflicts, particularly when few data are available, as is often the case in a phase I trial?
\end{itemize}
These questions will motivate the methodology developed in Sections \ref{sec:BayesMod} and \ref{sec:mixprior}. 
In what follows, we will define the MTD as the dose associated with a target DLT risk of $\Gamma$.

\section{Representing animal data in a prior for dose-toxicity parameters}
\label{sec:BayesMod}

Let $\mathcal{D} = \{d_1, \dots, d_{\mathcal{I}}; d_i < d_\ell \text{ for } 1\leq i < \ell \leq \mathcal{I} \}$ contain all doses available for evaluation. Furthermore, let $n_i$ denote the number of patients treated with dose $d_i$, of whom $r_i$ experience a DLT. Several authors \citep{BioPS:Whitehead1998, SIM:Neuenschwander2008, SiM:Zheng2018} have used a two-parameter BLRM to guide the dose escalation procedure in a phase I trial. We follow them to assume that DLT risk increases monotonically with dose and that this relationship can be described as:

\begin{equation}
\label{eq:dose-tox}
\begin{split}
r_i | p_i, n_i &\sim \text{Binomial}(p_i, n_i),  \quad \text{for } i = 1, \dots, \mathcal{I}, \\
\text{logit}(p_i) &= \theta_1 + \exp(\theta_2)\log(d_i/d_\text{Ref}),
\end{split}
\end{equation}

\noindent where $d_\text{Ref}$ is a pre-defined reference dose drawn from $\mathcal{D}$ and $\theta_1$ is the log-odds of toxicity on $d_\text{Ref}$. Model \eqref{eq:dose-tox} is parameterised differently to the two-parameter logistic regression model used by \citet{AACR:Sessa2013} who specify their model with intercept $\log(\theta_1)$ and slope $\theta_2$, constraining $\theta_1, \theta_2>0$ \citep{SIM:Neuenschwander2008}.
Our model ensures monotonicity and allows $\theta_1, \theta_2$ to take any real value. 
In what follows, we will focus on estimating the parameter vector $\boldsymbol{\theta}=(\theta_1, \theta_2)$ using all available animal and human data. \\

Suppose that preclinical data are available from one animal study performed in a single species thought to be relevant for predicting DLT risks in humans. The monotonicity assumption should remain applicable to animal data. To this end, we assume that the crude estimate of DLT risk on the lower animal dose does not exceed that on the higher dose. At a minimum, animal data must be available on two doses and at least one toxicity must have been observed on the higher dose. 
Indexing animal doses by $j = 0, -1$, if $t_{\mathcal{A}j}$ animals out of $(t_{\mathcal{A}j} + \nu_{\mathcal{A}j})$ on dose $d_{\mathcal{A}j}$ experienced a toxicity, 
we can represent the animal data by $\boldsymbol{x}_\mathcal{A}= \{(d_{\mathcal{A}j}, t_{\mathcal{A}j}, v_{\mathcal{A}j}); j = -1, 0 \}$. We follow \citet{SiM:Zheng2018} and use allometric scaling on the basis of body surface area \citep{FDA:FIH2005, BPH:Sharma2009} to
translate the animal dose-toxicity curve onto an equivalent human dosing scale, assuming differences in DLT risk between an animal species and humans, given the same dose, can be largely explained by differences in size.
In this way, we can identify doses $d_{-1}$ and $d_0$ such that for $j=0, -1$, the risk of DLT in humans given dose $d_j\, j = 0, -1$ is anticipated to be similar to the risk of a DLT in animals given dose $d_{\mathcal{A}j}$.
Doses $d_{-1}$ and $d_0$ are not necessarily contained in $\mathcal{D}$. For $j = -1, 0$, we stipulate an independent prior $p_j \sim$ Beta($t_{\mathcal{A}j}, v_{\mathcal{A}j}$) for the DLT risk in humans given dose $d_j$, which has effective sample size (ESS) equal to ($t_{\mathcal{A}j} + v_{\mathcal{A}j}$) \citep{BIOM:Morita2008}. 
In this way, the animal data on their original doses $d_{\mathcal{A}j}$ are taken to represent pseudo-observations of ($t_{\mathcal{A}j} + v_{\mathcal{A}j}$) patients on dose $d_j$.  \\

Using these independent beta priors for $p_{-1}$ and $p_0$, and assuming the dose-toxicity relationship follows Model \eqref{eq:dose-tox}, we can apply a Jacobian transformation to obtain the joint prior probability density function (pdf) of $\theta_2$ and $p_i$, for $1 \leq i \leq \mathcal{I}$, as:

\begin{equation}
\label{eq:jointPrior}
g_i(p_i, \, \theta_2 \mid \boldsymbol{x}_\mathcal{A}) = \frac{1}{p_i(1-p_i)} \exp(\theta_2)\left| \log \left( \frac{d_{-1}}{d_0} \right) \right|  \times \prod_{j=-1}^0 \frac{[1+\exp(-z_{ji})]^{-t_{\mathcal{A} j}}[1+\exp(z_{ji})]^{-v_{\mathcal{A} j}}}{B(t_{\mathcal{A} j}, v_{\mathcal{A} j})},
\end{equation}

\noindent where $z_{ji}={\rm logit} ( p_i) +\exp(\theta_2)\log(d_j/d_{\rm Ref})$, for $j = -1, 0$; and $B(a, b)$ is the beta function evaluated at $(a, b)$. A detailed derivation of $g_i(p_i, \, \theta_2 \mid \boldsymbol{x}_\mathcal{A})$ is provided in Appendix A. The marginal prior pdf of $p_i$ is given by

\begin{equation}
\label{eq:margiPrior}
f_i(p_i \mid \boldsymbol{x}_\mathcal{A}) = \int_{-\infty}^{+\infty} g_i(p_i, \, \theta_2 \mid \boldsymbol{x}_\mathcal{A}) {\rm d} \theta_2 \quad \text{for } 0 < {p_i} < 1,
\end{equation}

\noindent and the prior cumulative distribution function (cdf) for $p_i$ evaluated at $p_i = p$ is

\begin{equation}
\label{eq:pi_cdf}
F_i(p \mid \boldsymbol{x}_\mathcal{A}) = \int_0^p f_i(p_i \mid \boldsymbol{x}_\mathcal{A}){\rm d} p_i = \int_0^p \int_{-\infty}^\infty g_i(p_i,\, \theta_2 \mid \boldsymbol{x}_\mathcal{A}) {\rm d}\theta_2 {\rm d} p_i \quad \text{for } 0 < p < 1.
\end{equation}

We outline below how we can represent the preclinical information on toxicity probabilities, $p_1, \ldots, p_{\mathcal{I}}$, by a bivariate normal prior for $\boldsymbol{\theta}$.
The general idea of finding an approximate prior for $\boldsymbol{\theta}$ by matching percentiles of the marginal priors for $p_1, \ldots, p_{\mathcal{I}}$ is due to \citet{SIM:Neuenschwander2008}.

\begin{itemize}
\item[(i)]  For each $i = 1, \ldots, \mathcal{I}$, we summarise the prior for $p_i$ given in \eqref{eq:margiPrior} by $K$ percentiles. We recommend using three (or more) percentiles. In our illustrative examples, we find the median, together with the 2.5th and 97.5th percentiles. The $k$-th ($0<k\leq 100$) percentile of $p_i$, denoted by $q_{ik}$, is found as the solution to:
$$
\text{Pr}(p_i \leq q_{ik} \mid \boldsymbol{x}_\mathcal{A}) = \int_0^{q_{ik}}\int_{-\infty}^\infty g_i(p_i,\, \theta_2 \mid \boldsymbol{x}_\mathcal{A}) {\rm d}\theta_2 {\rm d} p_i = k.
$$
\item[(ii)] To obtain a bivariate normal prior $\boldsymbol{\theta} \sim {\rm BVN}(\boldsymbol{\mu}, \, \Sigma)$ that is consistent with the $K$ percentiles found in step (i), we use a modified quasi-Newton algorithm \citep{SIAM:Byrd1995}. With constraints set to the correlation of $\theta_1$ and $\theta_2$, the algorithm searches over configurations of $\boldsymbol{\mu}$ and $\Sigma$ to find the one minimising the total absolute distance between percentiles of the fitted marginal priors for the $p_i$, denoted by $q_{11}', \dots, q_{1K}', \dots, q_{\mathcal{I}1}', \dots, q_{\mathcal{I}K}'$, which are analytically available (see Appendix B), and the orginal percentiles calculated in step (i). This distance measure is formally defined as:
$$
\delta = \sum_{i=1}^{\mathcal{I}} \sum_{k=1}^K|q_{ik} - q_{ik}'|.
$$
\end{itemize}

Hereafter, we denote the pdf of the bivariate normal prior for $\boldsymbol{\theta}$ as $\pi_0(\boldsymbol{\theta} | \boldsymbol{x}_\mathcal{A})$. 
The above approach for deriving $\pi_0(\boldsymbol{\theta} | \boldsymbol{x}_\mathcal{A})$ assumed we had animal data on two doses. If instead we had data on more than two doses, we would use these to first derive independent beta priors for the DLT risk on each corresponding human dose; calculate the percentiles $q_{ik}$ of each beta prior; and then follow steps (i) and (ii) to find a bivariate normal approximation to the prior for $\boldsymbol{\theta}$. 
\\

\begin{figure}
\centering
\captionsetup{font=scriptsize}
\includegraphics[width=0.68\linewidth]{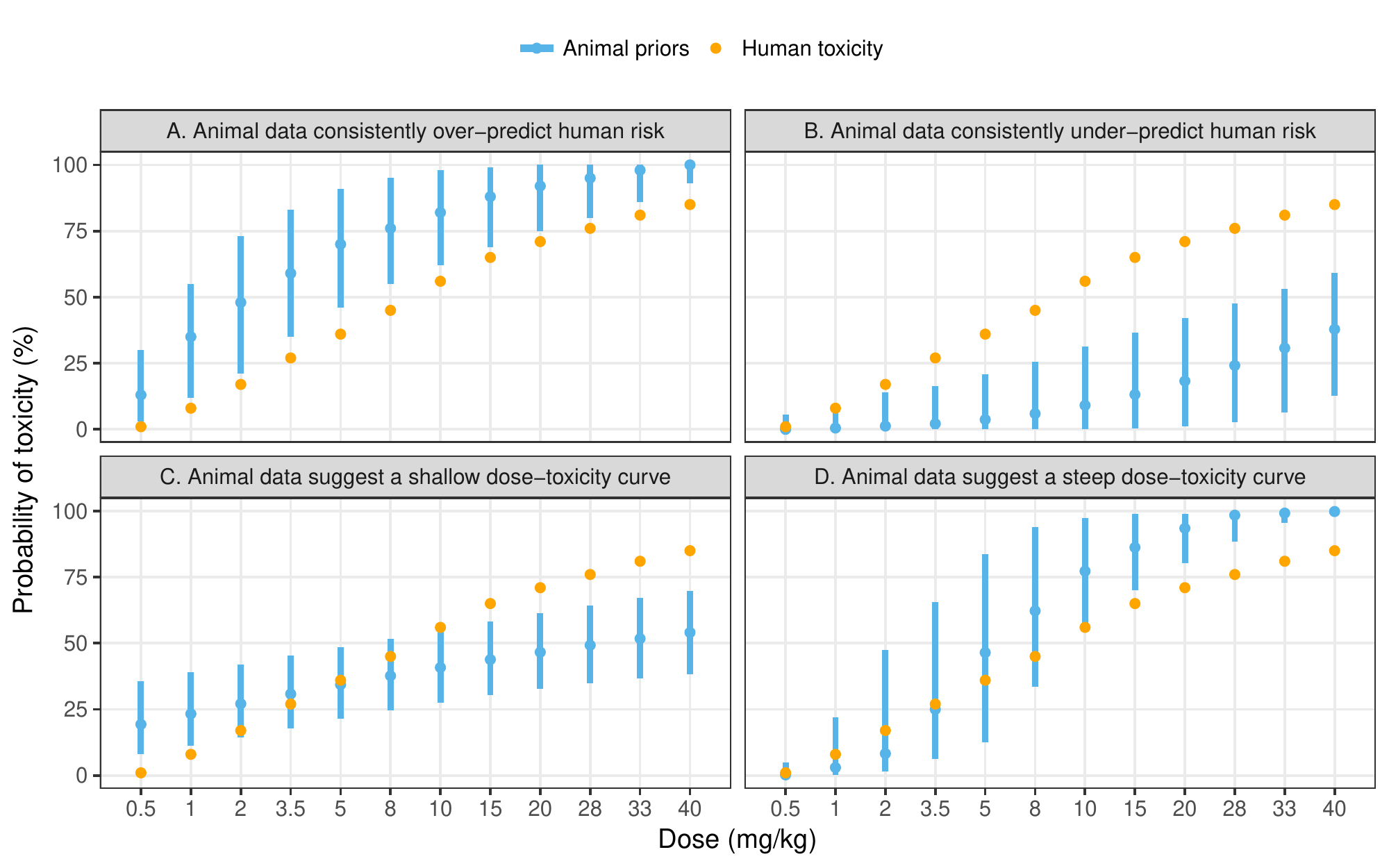}
\caption{Potential conflicts between the true (unknown) human toxicity and the priors obtained based on animal data alone.
On each dose, the orange point represents the true DLT risk in humans; the blue point and line represent the prior median and 95\% credible interval for human toxicity.}
\label{fig:AniPrior}
\end{figure}

It is possible that human data from the phase I clinical trial may conflict with what was expected based on $\pi_0(\boldsymbol{\theta} | \boldsymbol{x}_\mathcal{A})$. Figure \ref{fig:AniPrior} illustrates four ways in which preclinical data could conflict with the true (unknown) dose-toxicity relationship in humans. Preclinical data can (A) consistently over-predict or (B) consistently under-predict the human DLT risk, or a mixture of (A) and (B), over the therapeutic interval where doses have a DLT risk close to the target of $\Gamma=0.25$. We want to leverage the preclinical data to support inferences about $\boldsymbol{\theta}$ when the animal data are highly predictive of human data and quickly discount them otherwise. Another consideration is that the consequence of the prior-data conflict shown in Figure \ref{fig:AniPrior}A may be quite different to that of the conflict shown in Figure \ref{fig:AniPrior}B. This is because leveraging preclinical data to inform dose recommendations when they under-predict the human DLT risk may lead to overdosing patients and placing them at excessive risk.  \\

[Please insert Figure \ref{fig:AniPrior} about here.] \\

To facilitate robust borrowing of information across species, we implement Bayesian dose-escalation procedures with a robust mixture prior for $\boldsymbol{\theta}$, which is a weighted average of an informative component, $\pi_0(\boldsymbol{\theta} | \boldsymbol{x}_\mathcal{A})$, and a weakly informative component, denoted by $m_0(\boldsymbol{\theta})$. Robust mixture priors have been proposed to leverage historical controls in a new clinical trial \citep{BIOM:Heinz2014} and to extrapolate adult data to paediatrics \citep{SMMR:Wadsworth2016, PS:Gamalo-Siebers2017}. Denoting the prior weight attributed to the informative component by $w$, a mixture prior can be written as
\begin{equation}
\label{eq:mixp}
\mu_0(\boldsymbol{\theta} \mid \boldsymbol{x}_\mathcal{A}) = w \cdot \pi_0(\boldsymbol{\theta} \mid \boldsymbol{x}_\mathcal{A}) + (1-w) \cdot m_0(\boldsymbol{\theta}).
\end{equation}

Rather than defining $m_0(\boldsymbol{\theta})$ as a non-informative prior, we specify it so as to place probability mass on a wide range of plausible parameter values. We define $m_0(\boldsymbol{\theta})$ as the pdf of a bivariate normal distribution with correlation coefficient 0, setting $d_\text{Ref}$ as the dose in $\mathcal{D}$ most likely to be associated with target DLT risk $\Gamma$. We then stipulate $\theta_1 \sim N(m_{01}, \sigma_{01}^2)$, calibrating $m_{01}$ and $\sigma_{01}$ to ensure the median for the DLT risk on dose $d_\text{Ref}$ is $\Gamma$ and the 95\% credible interval is wide, say 0.01 to 0.95. A normal prior for $\theta_2 \sim N(m_{02}, \sigma_{02}^2)$ is calibrated to accommodate very flat to steep dose-toxicity curves in humans. The illustrative examples and simulation study described in Sections \ref{sec:examplePhI} and \ref{sec:sims} are performed defining $m_0(\boldsymbol{\theta})$ so as to place independent priors on $\theta_1$ and $\theta_2$ with $\theta_1 \sim N\left( \text{logit}(\Gamma), 2^2 \right)$ and $\theta_2 \sim N(0, 1^2)$.

\section{Leveraging animal data using a mixture prior with dynamic weights}
\label{sec:mixprior}

In this section, we propose analysing data from a ongoing phase I trial using a two-component mixture prior with dynamically chosen weights. Specifically, after completion of cohort $h$, we use Bayes' Theorem to update the prior distribution

\begin{equation}
\label{eq:mixPrior}
\mu^{(h)}_0(\boldsymbol{\theta} \mid \boldsymbol{x}_\mathcal{A})  = w^{(h)} \cdot \pi_0(\boldsymbol{\theta} \mid \boldsymbol{x}_\mathcal{A}) + (1-w^{(h)}) \cdot m_0(\boldsymbol{\theta}),
\end{equation}

\noindent with the trial data from cohorts $1, \dots, h$ to derive the posterior distribution for $\boldsymbol{\theta}$. The posterior formed by updating $\mu^{(h)}_0(\boldsymbol{\theta} \mid \boldsymbol{x}_\mathcal{A})$ with human data will also be a two-component mixture (O'Hagan and Forster, 2004), where the posterior weights will reflect the relative likelihood of the human data under each component of the prior. Note that $\pi_0(\boldsymbol{\theta} | \boldsymbol{x}_\mathcal{A})$ and $m_0(\boldsymbol{\theta})$ in \eqref{eq:mixPrior} are as defined in Section \ref{sec:BayesMod}, and do not depend on $h$. The cohort-specific prior mixture weight $w^{(h)} \in [0, 1]$ measures the similarity of the animal data and human data from cohorts $1, \ldots, h$. Weight $w^{(0)}$ will be based on expert prior opinion on the degree of similarity between the prior predictive distribution for the human dose-toxicity relationship formed from preclinical data, and the unknown human relationship. 
This will be useful for choosing a starting dose for the trial.
Rather than using a prior with the same prior mixture weight to analyse accumulating trial data, we revise this weight to encourage faster discounting of the preclinical information in the event of a prior-data conflict.
Figure \ref{fig:schem} is a schematic diagram of how a Bayesian dose-escalation procedure, driven by a mixture prior with dynamically chosen prior weights, proceeds.
In the following, we develop a Bayesian decision-theoretic  approach for calculating $w^{(h)}$ during an ongoing phase I dose-escalation trial.\\

[Please insert Figure \ref{fig:schem} about here.] \\

\subsection{Assessment of commensurability using a Bayesian decision-theoretic approach}
\label{sec:bdta}

Let $Y_i$ represent the binary DLT outcome of a new patient assigned dose $d_i\in \mathcal{D}$, such that $Y_i =1$ when a patient experiences a DLT, and $Y_i =0$ otherwise. Furthermore, let $y_i$ denote the realisation of $Y_i$. The prior predictive probability mass function of $Y_i$ given the animal data is written as:
\begin{equation}
\label{eq:ppred}
\text{Pr}(Y_i = {y_i} | \boldsymbol{x}_\mathcal{A}) = \int_0^1 p_i^{y_i}\cdot (1-p_i)^{1- {y_i}} f_i(p_i | \boldsymbol{x}_\mathcal{A}) {\rm d} p_i \qquad  \text{for } {y_i} \in \{0, 1 \},
\end{equation}

\noindent where $f_i(p_i | \boldsymbol{x}_\mathcal{A})$ is the pdf of the marginal prior for $p_i$, defined in \eqref{eq:margiPrior}, for $i = 1, \dots, I$. 
Before a patient is treated, we can use prior predictive distribution (\ref{eq:ppred}) to derive a prediction, $\eta_i$, for $Y_i$ \citep{SS:Vehtari2012}. At any point before or during the phase I trial, the same outcome is predicted for all patients administered a particular dose. After the patient's outcome is observed, we can compare $Y_i$ with $\eta_i$ and compute the utility of the prediction. Let $U(y = \ell, \eta = s)$ denote the utility of prediction $\eta = s$ for observation $y = \ell$, which can take values $u_{\ell s}$, for $\ell, s \in \{0, 1\}$. Utilities are dimensionless quantities, independent of dose, which lie between 0 and 1. Table \ref{tab:2by2rp} lists all possible configurations of the predicted and observed DLT outcomes, and corresponding utilities. We consider schemes setting $u_{00} = u_{11} = 1$, thus assigning correct predictions a utility of 1, and setting $u_{10}=0$, that is, assigning an incorrect prediction of no DLT zero utility. We explore setting $u_{01} = b$, with $0 < b < 1$, since incorrectly predicting a no DLT outcome as a DLT (unlike the reverse) will not undermine patients' safety but rather introduce unnecessary caution. Before $Y_i$ is observed, the optimal prediction for $Y_i$ based on the animal data is
\begin{equation}
\hat{\eta}_i = \arg \underset{\eta \in \{0, 1\}}{\max}  \sum_{y =0}^{1} U(y, \eta) \text{Pr}(Y_i = y),
\end{equation}

\noindent and we will take these optimal predictions for comparison with the observed human data.  \\

[Please insert Table \ref{tab:2by2rp} about here.] \\

Fouskakis and Draper (2002) suggest a metric that quantifies the discrepancy between predicted and observed data. We now adapt this idea to develop a measure of commensurability, which we re-evaluate on the completion of each cohort $h$ of the phase I trial and use to dynamically choose the prior mixture weights in \eqref{eq:mixPrior}. We measure the commensurability of the preclinical and human data by the average utility of the optimal predictions derived from animal data, taking averages across a suitable dosing range. We now describe in more detail the measure of commensurability.  \\

Suppose cohort $h \geq 1$ has just been treated and observed. We want to calculate the prior mixture weight $w^{(h)}$ that will define the mixture prior used to analyse the human data from cohorts 1, \ldots, $h$ (inclusive).
Let $\mathcal{D}^{(h)} \subseteq \mathcal{D}$ contain all the doses that have been tested by this stage. For each $d_i \in \mathcal{D}^{(h)}$, we summarise the total number of patients who have: a) received dose $d_i$ (in cohorts $1,\dots,h$); and b) were predicted toxicity outcome $\hat{\eta}_i = s$; and c) experienced outcome $y = \ell$ as $n_{i, \ell s}^{(h)}$. We then define the predictive utility of the preclinical information per administered dose $d_i \in \mathcal{D}^{(h)}$ as:

\begin{equation}
c_i^{(h)} = \frac{\sum_{\ell=0}^1 \sum_{s=0}^1 u_{\ell s}n_{i, \ell s}^{(h)}}{\sum_{\ell=0}^1 \sum_{s=0}^1 n_{i, \ell s}^{(h)}} \quad \text{for } d_i \in \mathcal{D}^{(h)}
\end{equation}

\noindent where the denominator is the maximum utility that would be achieved if all predictions were correct. Clearly $c_i^{(h)}$ only measures the predictive utility of the animal data based on outcomes from the first $h$ cohorts and on one dose. To measure the predictive utility of the animal data on all `interesting' doses at stages $h\geq 2$, we define $T^{(h)} \subseteq \mathcal{D}^{(h)}$ as the set of doses administered to cohorts $1, \dots, h$, which are not more than one dose level lower than the current posterior estimate of the target dose. Therefore, $T^{(h)}$ comprises tested doses in the neighbourhood (i.e. within one dose level) of the current MTD estimate, as well as all other tried doses which are estimated to have a DLT risk exceeding $\Gamma$. The current estimate of the MTD, the same as the dose given to cohort $h$, is based on the posterior that is updated from the mixture prior $\mu^{(h-1)}_0(\boldsymbol{\theta}| \boldsymbol{x}_\mathcal{A})$ with human data from cohorts $1, \ldots, (h-1)$. We measure the commensurability of animal and human data by averaging $c_i^{(h)}$ across $d_i \in T^{(h)}$, that is,

\begin{equation}
\kappa^{(h)} = \frac{1}{|T^{(h)}|} \sum_{\{i: d_i \in T^{(h)}\}} c_i^{(h)},
\end{equation}

\noindent where $|T^{(h)}|$ is the number of doses in $T^{(h)}$. Averaging across $T^{(h)}$ rather than $\mathcal{D}^{(h)}$ can be thought of as an {\it ad-hoc} constraint, but one which we apply to avoid measures of $\kappa^{(h)}$ being artificially biased upwards due to some very safe doses being included by conservative algorithms.
To explain why this bias might arise, we note that prior to starting the trial, we usually define $\mathcal{D}$ to include doses thought to be very safe based on the preclinical data. At these safe doses, even if there were differences between the DLT risk predicted from animal data and the true human risk, these differences would likely be small (in absolute terms) and unlikely to manifest themselves in discrepancies between the optimal predictions and observed human DLT outcomes. Therefore, the value of $c_i^{(h)}$ for these low doses would be expected to be close to 1, even in the presence of divergent dose-toxicity relationships in animals and humans. The quantity $\kappa^{(h)}$ is used to determine the prior mixture weight in \eqref{eq:mixPrior} and thus $\mu_0^{(h)}(\boldsymbol{\theta} | \boldsymbol{x}_\mathcal{A})$, which will be used to analyse trial data from cohorts 1, \ldots, $h$.

\subsection{Choosing an appropriate tuning parameter}

When using $\kappa^{(h)}$ to calculate $w^{(h)}$ in \eqref{eq:mixPrior}, we must bear in mind that $\kappa^{(h)}$ will be a noisy measure of the predictive utility of the animal data, particularly in the early stages of the phase I trial when few doses have been tried and few patients have been treated. As more data accrue, the assessment of predictive utility becomes more convincing. To reflect this, we would not suggest setting $w^{(h)} = \kappa^{(h)}$, but propose discounting $\kappa^{(h)}$ according to a power law relationship, with

\begin{equation}
w^{(h)} = \left\{ \kappa^{(h)} \right\} ^{\lambda^{(h)}} \quad \text{for } h  =1, \dots, H
\end{equation}

\noindent where $\lambda^{(h)}$ is a time-dependent tuning parameter and $\lambda^{(1)}, \ldots, \lambda^{(H)} \geq 1$ are a decreasing sequence of powers. We consider two approaches for defining each $\lambda^{(h)}$. For any choice of $\lambda^{(h)}$, a weight $w^{(h)} = 1$ will be attributed to the informative prior component in \eqref{eq:mixPrior}, if preclinical data correctly predict all DLT outcomes in humans in the first $h$ cohorts, since this leads to $\kappa^{(h)}=1$ according to our decision-theoretic approach.  
A simple proposal is to relate $\lambda^{(h)}$ with the trial information time:

\begin{equation}
\lambda^{(h)} = \sqrt{N/n^{(h)}},
\label{eq:lambdaThall}
\end{equation}

\noindent where $N$ and $n^{(h)}$ are the maximum sample size and the number of patients recruited in the first $h$ cohorts, respectively. 
It follows that by definition $\lambda^{(H)}=1$ on completion of the phase I trial.
A similar power law function, relating the exponent with the trial information time, was used in a different context by \citet{EJC:Thall2007} for response adaptive randomisation.  \\

Suppose $d_{i^{\star}} \in \mathcal{D}$ was recommended for use in cohort $h$. We could also consider defining $\lambda^{(h)}$ to capture how noisy our estimate of the predictive utility of the animal data for patients receiving dose $d_{i^{\star}}$ is. Let $\hat{\eta}_{i^{\star}}$ denote the optimal prediction from animal data for the outcome of a patient on dose $d_{i^{\star}}$, and $H$ be the maximum number of cohorts planned for the phase I trial. 
Before observing cohort $h$, both $c_{i^{\star}}^{(h)}$ and $c_{i^{\star}}^{(H)}$ are random variables. The standard deviation of $c_{i^{\star}}^{(h)}$, written as $\sigma\{c_{i^{\star}}^{(h)}\}$, can be calculated analytically. To do this, we note that given a prior prediction $\hat{\eta}_{i^{\star}}=0$ ($\hat{\eta}_{i^{\star}}=1$), the utility can take values of either $u_{00}$ or $u_{10}$ ($u_{01}$ or $u_{11}$) with probabilities ${\rm Pr}(Y_{i^{\star}} = 0)$ and ${\rm Pr}(Y_{i^{\star}} = 1)$, which can be set equal to the posterior modal estimates, calculated based on the posterior updating $\mu_0^{(h-1)}(\boldsymbol{\theta} | \boldsymbol{x}_\mathcal{A})$ with human data from cohorts 1, \ldots, $(h-1)$. To compute $\sigma\{c_{i^{\star}}^{(1)}\}$, ${\rm Pr}(Y_{i^{\star}} = 0)$ and ${\rm Pr}(Y_{i^{\star}} = 1)$ are set equal to the prior estimates based on animal data alone, that is, from $\mu_0^{(0)}(\boldsymbol{\theta} | \boldsymbol{x}_\mathcal{A})$ with $w^{(0)}=1$.
Computing the standard deviation of $c_{i^{\star}}^{(H)}$, written as $\sigma\{c_{i^{\star}}^{(H)}\}$, is less straightforward.
We can estimate it using simulation, setting $\text{Pr}(Y_{i^{\star}}=1)$ equal to the latest posterior modal estimate of the DLT risk on dose $d_{i^{\star}}$.
We simulate the binary outcomes of patients in cohorts $h, \ldots, H$, assuming all these patients receive dose $d_{i^{\star}}$, to compute $c_{i^{\star}}^{(H)}$. We repeat this process 5,000 times and calculate the empirical standard deviation of the simulated values of $c_{i^{\star}}^{(H)}$. In this way, $\lambda^{(h)}$ is defined as:

\begin{equation}
\lambda^{(h)} = \frac{\sigma \{ c_{i^{\star}}^{(h)} \} }{\sigma \{ c_{i^{\star}}^{(H)} \} },
\label{eq:lambda}
\end{equation}

\noindent which reflects how noisy $c_{i^{\star}}^{(h)}$ is compared with $c_{i^{\star}}^{(H)}$. This definition of $\lambda^{(h)}$ implicitly reflects the trial information time, since $\lambda^{(h)}$ tends to be large in early stages of the trial and converges towards 1 by the end. In Section \ref{sec:sims}, we will compare the operating characteristics of dose-escalation procedures using $\lambda^{(h)}$ as defined in (\ref{eq:lambdaThall}) and (\ref{eq:lambda}).

\section{Design and analysis of the example trial incorporating animal data}
\label{sec:examplePhI}

In this section, we illustrate how the proposed approach can be used to incorporate animal data into a phase I dose-escalation study. Here, we define the target DLT risk as $\Gamma = 0.25$.

\subsection{Prior distributions based on preclinical information}
\label{sec:aniPrior}

We return to the AUY922 example described in Section \ref{sec:motivate}. For illustrative purposes, we suppose that 0.1 and 2.7 mg/kg have each been tested in 30 dogs, of which 1 and 17 experienced DLTs, respectively. Using allometric scaling to standardise body weight (BW) by body surface area (BSA), one can calculate the human doses equivalent to 0.1 and 2.7 mg/kg in dogs as:
\begin{equation*}
{\rm Equivalent\; human \; dose \; (mg/m}^2)  =  {\rm Animal \; dose\; (mg/kg)}\times \left(\frac{\rm BW}{\rm BSA}\right)_{\rm Animal},
\end{equation*}
\noindent where the right-hand side translation factor, defined by BW and BSA, appropriate for dogs is (10/0.5), as specified in the FDA draft guideline {\it Estimating the Maximum Safe Starting Dose in Initial Clinical Trials for Therapeutics in Adult Healthy Volunteers} (FDA, 2005). Thus, the equivalent human doses are 2 mg/m$^2$ (labelled as dose $d_{-1}$) and 54 mg/m$^2$ (labelled as dose $d_0$), with corresponding risks of toxicity $p_{-1}$ and $p_0$ in humans. We summarise the dog data as prior distributions $p_{-1} \sim \text{Beta}(1, 29)$ and $p_0 \sim \text{Beta}(17, 13)$.  \\

Recall that for the AUY922 phase I trial, $\mathcal{D} = \{2, 4, 8, 16, 22, 28, 40, 54, 70\}$ mg/m$^2$. The human equivalent doses for the available dog data happen to lie in $\mathcal{D}$. Assuming that the human dose-toxicity relationship is described by Model \eqref{eq:dose-tox}, we can follow steps (i) and (ii) in Section \ref{sec:BayesMod} to compute the 2.5th, 50th and 97.5th percentiles of the marginal prior distributions of $p_1, \ldots, p_9$ and search to find a bivariate normal approximation $\pi_0(\boldsymbol{\theta} \mid \boldsymbol{x}_\mathcal{A})$ which is given by:

\begin{equation}
\boldsymbol{\theta} \mid \boldsymbol{x}_\mathcal{A} \sim \text{BVN}\left( \begin{pmatrix}
-0.524  \\
\;\;\; 0.147
\end{pmatrix},
\begin{pmatrix}
\;\;\; 0.151 & -0.008 \\
-0.008 & \;\;\; 0.001
\end{pmatrix}
\right).
\label{eq:normApprox}
\end{equation}
\noindent Figure \ref{fig:aPriors}A compares the percentiles of the marginal priors for $p_i, \, i = 1, \dots, 9$, with the fitted percentiles corresponding to the bivariate normal distribution defined above.  \\

[Please insert Figure \ref{fig:aPriors} about here.] \\

From Figure \ref{fig:aPriors}A we see that our hypothetical data on 60 dogs suggest that doses 16 and 22 mg/m$^2$ have risk of toxicity in humans close to $\Gamma$. Figures \ref{fig:aPriors}B -- \ref{fig:aPriors}C summarise marginal prior distributions for $p_1, \dots, p_9$ calculated based on the bivariate normal approximation in \eqref{eq:normApprox}. Figure \ref{fig:aPriors}B summarises each marginal prior for $p_i, i = 1,\dots,9,$ by three interval probabilities: (i) the probability of underdosing, said to occur when $p_i\in [0, 0.16)$; (ii) the probability that $p_i$ lies in the target interval $[0.16, 0.33)$; and (iii) the probability of overdosing, said to occur when $p_i$ exceeds 0.33 \citep{SIM:Neuenschwander2008}. Figure \ref{fig:aPriors}C presents the prior probability density curves for the lowest doses in $\mathcal{D}$. In the illustrative examples, we set dose 4 mg/m$^2$ as the starting dose for the phase I trial as it appears to be safe with $\text{Pr}(p_2 < 0.1 \mid \boldsymbol{x}_\mathcal{A}) = 0.825$. \\

To evaluate the effective sample size (ESS) of the marginal prior for $p_i$, we approximate it by a Beta($a, b$) distribution matching the first and second moments. The ESS is then $(a+b)$. Table \ref{tab:pseudopts} lists prior ESSs of the human DLT risks for our hypothetical example. The preclinical data provide information on the DLT risks at low doses (2 to 8 mg/m$^2$) equivalent to what would have been obtained from, approximately, 23 -- 63 humans, and at moderate to high doses (16 to 70 mg/m$^2$) equivalent to that from 15 -- 17 humans. This prior may overwhelm the data from a small phase I trial. It is important to see whether our approach for dynamically leveraging animal data can lead to a robust dose-escalation procedure, particularly when there is a prior-data conflict. \\

[Please insert Table \ref{tab:pseudopts} about here.] \\

\subsection{Hypothetical data examples}
\label{sec:hypoPhI}

From Figure \ref{fig:aPriors}B, we see that ahead of the phase I trial, doses up to 16 mg/m$^2$ comply with the escalation criterion defined in (\ref{eq:overdose}) if probabilities are based on the prior in \eqref{eq:mixPrior} with $h=0$ and $w^{(0)} = 1$. 
We analyse data from the first human cohort, for which dose 4 mg/m$^2$ has been chosen based on animal data alone.
Dynamic updating of the prior mixture weight commences after completion of the first cohort onwards, when the accumulating human toxicity data are compared with the prior predictions obtained using animal data.
We illustrate how to implement the Bayesian dose-escalation procedure driven by $\mu_0^{(h)}(\boldsymbol{\theta} | \boldsymbol{x}_\mathcal{A})$ with the following utility values. A utility of $u_{01} = 0.6$ is considered for quick discounting of animal data which predict a no-DLT as DLT, and a zero utility of $u_{10}=0$ for penalising incorrect predictions of a DLT as no-DLT. Based on the animal data in Figure \ref{fig:aPriors}, these utilities result in optimal prediction of no-DLT ($\hat{\eta} = 0$) for patients receiving doses 2, 4, 8 or 16 mg/m$^2$, and a DLT ($\hat{\eta} = 1$) for  patients receiving doses 22, 28, 40, 54 or 70 mg/m$^2$.  \\

We simulated three data examples, where there is (i) a negligible prior-data conflict; (ii) a conflict where the dog data under-estimate the DLT risk in humans; and (iii) a conflict where the dog data over-estimate the DLT risk in humans. 
The dynamic mixture prior $\mu_0^{(h)}(\boldsymbol{\theta} \mid \boldsymbol{x}_\mathcal{A})$ was used for leveraging animal data to analyse the phase I dose-escalation trial, with the prior mixture weight empirically determined by the assessed commensurability $\kappa^{(h)}$ and the tuning parameter $\lambda^{(h)}$ defined in \eqref{eq:lambda}.
Patients were recruited in cohorts of size three and each hypothetical trial can recruit a maximum of 11 cohorts. 
Once data are available from cohorts $1, \ldots, h$, the prior is updated using Bayes' Theorem to obtain the posterior:
\begin{equation}
\label{eq:mixPosterior}
\mu^{(h)}(\boldsymbol{\theta} \mid \boldsymbol{x}_\mathcal{A}, \boldsymbol{x}_\mathcal{H}^{(h)}) = w^{(h)}_\ast \cdot \pi^{(h)}(\boldsymbol{\theta} \mid \boldsymbol{x}_\mathcal{A}, \boldsymbol{x}_\mathcal{H}^{(h)}) + (1-w^{(h)}_\ast) \cdot m^{(h)}(\boldsymbol{\theta} \mid \boldsymbol{x}_\mathcal{H}^{(h)}),
\end{equation}

\noindent where $\boldsymbol{x}_\mathcal{H}^{(h)}$ denotes the human data from the first $h$ cohorts and $w^{(h)}_\ast$ denotes the weight attributed to the component $\pi^{(h)}(\boldsymbol{\theta} \mid \boldsymbol{x}_\mathcal{A}, \boldsymbol{x}_\mathcal{H}^{(h)})$, updated from $\pi_0(\boldsymbol{\theta} \mid \boldsymbol{x}_\mathcal{A})$. 
Dose recommendation for cohort ($h+1$) is based on $\mu^{(h)}(\boldsymbol{\theta} \mid \boldsymbol{x}_\mathcal{A}, \boldsymbol{x}_\mathcal{H}^{(h)})$. 
We fit our Bayesian model using Markov chain Monte Carlo. The OpenBUGS \citep{SiM:Lunn2009} code for implementation is available from the publisher's website. \\

[Please insert Figure \ref{fig:dataSc} about here.] \\

Figure \ref{fig:dataSc} presents the dosing trajectories and evolution of the prior mixture weight $w^{(h)}$ for the three data examples. Readers can find the exact values of  $w^{(h)}$, together with the arguments $\lambda^{(h)}$ and $\kappa^{(h)}$, in Table S1 of the Web-based Supplementary Materials. Looking at data examples 1 and 3, we see that the disagreement between predictions and observed data may take a few cohorts to emerge, even when there is a conflict between the true human dose-toxicity curve and predictions based on the animal data, as was noted in Section \ref{sec:bdta}.
The weight allocated to $\pi_0(\boldsymbol{\theta} \mid \boldsymbol{x}_\mathcal{A})$ in $\mu_0^{(h)}(\boldsymbol{\theta} | \boldsymbol{x}_\mathcal{A})$ decreases immediately from 1, after the first disagreement has been observed, for example, in the first cohort of data example 2. 
Looking across the three data examples, the prior mixture weight typically plateaus or increases slightly over the last couple of cohorts. This is because the predictive utility $\kappa^{(h)}$ that measures the commensurability of the human and animal data is unlikely to vary substantially over the final stages of the trial, while the tuning parameter $\lambda^{(h)}$ reduces to 1 as the trial nears completion.  \\

We focus on data examples 2 and 3 to learn how the prior mixture weight evolves in the presence of a prior-data conflict. In data example 2, $w^{(1)}$ drops from 1 to 0.26, as dose 4 mg/m$^2$ yielded a DLT unexpectedly. It later increases to $w^{(3)}= 0.77$ due to correct predictions of outcomes in cohorts 2 and 3 on the two lowest doses. When all three patients in cohort $h=4$ administered 8 mg/m$^2$ experience DLT outcomes which the animal data incorrectly predict as no DLTs, $w^{(4)}$ decreases substantially to 0.08.
Data example 3 shows how the procedure reacts to a prior-data conflict when the DLT risks in humans are much lower than predicted by the animal data. 
In cohorts 4 and 5 assigned 22 mg/m$^2$ and 28 mg/m$^2$, respectively, no DLTs are observed which contradicts the prior predictions based on animal data. Consequently, the prior weights used for analysis drop to $w^{(4)} = 0.533$ and $w^{(5)}=0.250$.
The reduced amount of borrowing results in an escalation to dose 54 mg/m$^2$, at which one out of three patients experiences a DLT.
Results comparing dose-escalation trajectories when using a weakly-informative prior versus $\pi_0(\boldsymbol{\theta | \boldsymbol{x}_\mathcal{A}})$ are shown in Figure S1 of the Web-based Supplementary Materials.

\subsection{Specifying a run-in period}
\label{sec:run-in}

The proposed approach may be thought of as counterintuitive: we often assign full weight to animal data in the early stages of a trial when few human data are available to assess commensurability. 
This is particularly true since we know the agreement between prior predictions, based on animal data alone, and the observed human data may be an artefact of starting the trial with very safe doses rather than a reflection of genuine agreement. If the true dose-toxicity relationship in humans has a very steep slope, placing full weight on the animal prior could lead to overly aggressive escalation decisions and unexpected DLTs.   \\

To this end, we consider a constrained version of our approach with a run-in period. We set the prior mixture weight equal to 0 until the first discrepancy is observed in cohort $h^\star$. From cohort ($h^\star$+1) onwards, accumulated human data are then analysed with a mixture weight calculated as described in Sections \ref{sec:BayesMod} and \ref{sec:mixprior}. Under this scheme, the dose-escalation trial begins basing interim decisions on a weakly-informative prior $m_0(\boldsymbol{\theta})$. To reach a fair conclusion about the impact of having a run-in period, we present three new data examples in Figure \ref{fig:dataScRunIn} where observations are sampled without replacement from the same simulated datasets used to create the corresponding data examples in Figure \ref{fig:dataSc}. For instance, focusing on
data example 1, the first patient assigned 28 mg/m$^2$ in Figure \ref{fig:dataSc} will have the same simulated toxicity outcome as the first patient assigned 28 mg/m$^2$ in Figure \ref{fig:dataScRunIn}. \\

[Please insert Figure \ref{fig:dataScRunIn} about here.] \\

Implementing the run-in period has no impact on data example 2, since the first conflict between the animal predictions and the human outcomes occurred in cohort 1. For data examples 1 and 3, we notice that dose escalation decisions tend to be less conservative when using the run-in period, especially in the early stages of the simulated trials, leading to different interim dose recommendations, as shown in Figure \ref{fig:dataScRunIn}. By the end of each simulated trial implementing the constrained version of our approach, the prior mixture weights $w^{(11)}$ are equal to the values obtained when we implement the unconstrained version.
This is due to (i) our way of simulating the human toxicity data, which ensures the same likelihood of the human toxicity across our illustrative examples presented in Figures \ref{fig:dataSc} and \ref{fig:dataScRunIn}, and (ii) the fact that the prior mixture weight is dynamically determined based on the compatibility of preclinical and clinical trial data.

\section{Simulation study}
\label{sec:sims}

\subsection{Basic settings}

We continue with our motivating example from Sections \ref{sec:motivate} and \ref{sec:examplePhI}, and evaluate the operating characteristics of dose-escalation trials conducted using unconstrained or constrained version of the proposed approach. Comparisons are made with Bayesian procedures basing inferences on mixture priors with fixed weights. We consider:

\begin{itemize}
\item Procedure A: Bayesian mixture prior with dynamic decision-theoretic weights; no run-in period
\item Procedure B: Bayesian mixture prior with dynamic decision-theoretic weights; run-in period
\item Procedure C: Bayesian mixture prior with $w^{(h)} = 0.5$ for all $h$; no run-in period
\item Procedure D: Bayesian informative prior, that is, $w^{(h)} = 1$ for all $h$; no run-in period
\item Procedure E: Bayesian weakly-informative prior, that is, $w^{(h)} = 0$ for all $h$
\end{itemize}

\noindent Procedures A and B are implemented using \eqref{eq:lambda} to calculate the tuning parameter $\lambda^{(h)}$. We use a fixed weight of 0.5 for Procedure C to ensure the mixture prior has heavy enough tails to adequately deal with most prior-data conflicts. We use the optimal non-parametric design \citep{Bios:OQuigley2002} as a benchmark for comparison.   \\

We assume the hypothetical data on 60 dogs described in Section \ref{sec:aniPrior} are available prior to the phase I trial. 
To evaluate how rapidly procedures react to a prior-data conflict, we simulate small phase I trials recruiting up to seven cohorts of three patients, with doses in $\mathcal{D}$ available for evaluation.
As explained in Section \ref{sec:aniPrior}, patients in cohort 1 are assigned dose 4 mg/m$^2$.
After completion of cohort $(h-1)$, the dose for cohort $h$ is selected according to a variant of criterion (\ref{eq:overdose}) such that:
$$
d_\text{sel}^{(h)} = \max\{d_i \in \mathcal{D}: \text{Pr}(p_i \geq 0.33 \mid \boldsymbol{x}_\mathcal{A}, \boldsymbol{x}_\mathcal{H}^{(h-1)}) \leq 0.25\},
$$
with the same constraint on a maximum two-fold escalation in dose. Trials end either when all 21 patients have been treated and observed, or after any stage $h$ if $\text{Pr}(p_1 \geq 0.33 \mid \boldsymbol{x}_\mathcal{A}, \boldsymbol{x}_\mathcal{H}^{(h)})>0.25$. These two subsets of simulated trials will be referred to as {\sl completed} or {\sl stopped early} trials, respectively. \\

[Please insert Table \ref{tab:base} about here.] \\

Phase I dose-escalation trials are simulated under the eight human toxicity scenarios listed in Table \ref{tab:base}. In Scenario 3 (a prior-data consistency scenario), the true DLT risks are consistent with the prior medians obtained from the dog data, which are illustrated in Figure \ref{fig:aPriors}A and derived assuming that the dose-toxicity relationship follows a logistic regression model.  With this exception, in none of the other scenarios were the human DLT risks derived from a logistic regression model. For Procedures A -- E, estimates of operating characteristics are based on 1000 simulated trials per toxicity scenario.  \\

Let $\tilde{p}_i$ denote the posterior median DLT risk for $d_i \in \mathcal{D}$ recorded on completion of the trial. At the end of a {\sl completed} trial, we declare the target dose to be

$$
\hat{d}_\text{M} = \arg \underset{d_i \in \mathcal{D}_c}{\min} |\tilde{p}_i - 0.25 |,
$$
where $\mathcal{D}_c \subseteq \mathcal{D}$ comprises all doses that have been administered to humans during the trial and satisfy safety criterion \eqref{eq:overdose}. For each scenario, we report the percentage of simulated trials {\sl stopped early} for safety without selecting a target dose, and the percentage of simulated trials which are completed declaring dose $d_i$ as the target dose, for $i = 1, \ldots, 9$. 
We also report the average number of patients allocated to each dose across the 1000 simulated trials.

\subsection{Results}
\label{sec:simResults}

Versions of Procedures A and B using different utility values are studied;
Figure \ref{fig:PhI} visualises the subset of simulation results when $u_{01}=0.6$. Complete numerical results (including those when we set $u_{01}=0.2$) are given in Table S3 of the Supplementary Materials.  \\

[Please insert Figure \ref{fig:PhI} about here.] \\

Focusing on Procedures A and B, we can draw conclusions about the value of incorporating a run-in period. We see that Procedures A and B perform similarly in almost all scenarios except Scenario 8, where dog data over-predict the human DLT risk: the drug is in fact very safe in humans with the highest dose being the MTD. 
Procedure A, which leverages animal data from the outset of the trial, tends to be more cautious in the early stages, with more patients treated at low, very safe doses. Procedure A is slow to discount the animal data because of correct prior predictions at low doses up to 16 mg/m$^2$. In contrast, Procedure B (with a run-in period) allows for a quicker escalation in this scenario, since animal data are leveraged only after the first disagreement between the prior prediction and human response. 
By this point, we had typically escalated up to higher doses, where the prior-data conflict was more likely to manifest itself in the observed data. 
This explains why, in Panel (ii), on average only 1.3 out of 21 patients are allocated to 70 mg/m$^2$ by Procedure A, while slightly more (2.4 out of 21) by Procedure B. Panel (i) shows that the PCS for Procedure B in Scenario 8 is 33.8\% compared with 20.3\% for Procedure A. A similar line of reasoning explains differences between Procedures A and B in Scenario 6.  \\

Procedure E selects dose 70 mg/m$^2$ as the MTD for 58.0\% of the simulated trials in Scenario 8. This reveals a weakness of the proposed approach in situations where there are differences between the prior dose-toxicity relationship and the true curve in humans, but these are too small to result in discrepancies between the prior predictions and the human outcomes. Setting a run-in period (Procedure B) improves operating characteristics, as it enables us to escalate up to higher doses where there is a greater chance of observing a disagreement between animal predictions and human responses. \\

Comparing Procedures B and C, we see that the former leads to an increase in PCS from 38.1\% to 48.7\% in Scenario 3, when animal and human data are commensurate. In other scenarios, both procedures have similar operating characteristics. In results not shown here, we found that in scenarios where there is a prior-data conflict, Procedure B achieves higher PCS than versions of Procedure C with fixed prior mixture weights set equal to a constant between 0.5 and 1. Comparing Procedures B, D and E, we see that Procedure B offers a satisfactory compromise between full incorporation of animal data and no borrowing at all. While Procedure D achieves maximum PCS in Scenarios 2 -- 4 when animal and human data are commensurate, it has poor properties in Scenarios 6 and 8 when there is a prior-data conflict. The situation is reversed for Procedure E. Overall, Procedure B maintains reasonable operating characteristics over a range of scenarios.  \\

Results for Procedures A and B with tuning parameter $\lambda^{(h)}$ calculated using \eqref{eq:lambdaThall}, as visualised by Figure S2 of the Supplementary Materials, are similar to what has been reported here.
We have also run simulations setting the maximum trial sample size as 33 (i.e., 11 patient cohorts) and 45 (i.e., 15 patients cohorts); see Figure S3 of the Supplementary Materials for the comparison between Procedures A -- E with 33 patients per simulated trial. Similar conclusions are drawn, with the exception that differences between Procedures A and B for the PCS in Scenario 8 diminish with increasing maximum trial sample size. This is because as the number of patients increases, the trial can escalate up to the highest dose when using Procedure A before the maximum sample size is reached. 
We also observe that a smaller utility such as $u_{01}=0.2$ is more influential on Procedure A than Procedure B. In particular, setting $u_{01}=0.2$ means the optimal prediction based on the same dog data are: no DLT for patients receiving doses up to 28 mg/m$^2$ but a DLT on doses 40, 54 and 70 mg/m$^2$. In Scenarios 6 and 8, low doses are very safe and would likely to yield a no DLT for patients receiving them. Using Procedure A, $\pi_0(\boldsymbol{\theta} | \boldsymbol{x}_\mathcal{A})$ tend to have a large $\kappa^{(h)}$ and thus a prior mixture weight $w^{(h)}$ very close to 1. Given the informativeness of $\pi_0(\boldsymbol{\theta} | \boldsymbol{x}_\mathcal{A})$ used in simulations, the posterior would recommend a dose most likely to be the MTD based on the dog data alone, which in fact is very safe and would yield $w^{(h+1)}$ again as nearly 1. Using Procedure B, however, the difference brought by the utility value of $u_{01}$ is not much, but a smaller $u_{01}$ generally leads to a reduced PCS in the same scenario.

\section{Discussion}
\label{sec:discuss}

The question of using historical data in a new clinical trial has been extensively discussed.
But little has been said about leveraging preclinical information in a phase I first-in-man trial, where there are unique considerations.
In this context, the challenge is to use the preclinical data to increase precision when there is a good degree of agreement between the preclinical information and human data, and otherwise to rapidly discount it. Detecting a prior-data conflict is made more challenging in a sequential trial with a small sample size.
Here, ``small'' is meant in relation to the prior ESS of the informative prior component based on animal data.  In this paper, we have proposed a solution to measure the commensurability of animal data (translated onto a human dosing scale) and human DLT outcomes observed during an ongoing phase I oncology trial. The degree of agreement is measured using a Bayesian decision-theoretic approach, quantifying the utility of optimal predictions based on animal data. Animal data leading to incorrect prior predictions will have a small predictive utility.  \\

So far we have assumed that animal data are available on two doses. After translation to an equivalent human dosing scale, prior opinions on the DLT risk in humans on these doses are represented by two Beta distributions. These prior opinions are then converted into a collection of marginal priors for the DLT risk on the human doses. To obtain a bivariate normal approximation to the marginal priors, we minimise an objective function that defines distances between the approximation and the marginal priors for $p_1, \dots, p_I$, as the sum of absolute distances between fitted and exact percentiles. One limitation of the current work is that we have not investigated the impact of using other distance measures. Since the use of another distance measure would likely impact on the eventual prior  component $\pi_0(\boldsymbol{\theta} | \boldsymbol{x}_\mathcal{A})$, this area deserves futher exploration. \\

In this paper, we have considered both unconstrained and constrained versions of the proposed Bayesian decision-theoretic approach for using animal data to drive a dose-escalation procedure. The constrained version features a run-in period, which allows animal data to be leveraged only after observing a discrepancy between a prediction and observed human responses. It results in acceptable trial operating characteristics for scenarios where the drug is much safer in humans than predicted. 
We acknowledge that alternative approaches could be used to define the length of the run-in period. For example, one reviewer pointed out that the run-in period could end once the ratio of the number of patients treated to the ESS of $\pi_0(\boldsymbol{\theta} | \boldsymbol{x}_\mathcal{A})$ exceeds a pre-specified threshold. 
We would also like to note that choosing appropriate utilities could be crucial for implementing the unconstrained version. Based on extensive numerical studies, we recommend setting $u_{01}$ as $0<b<1$ that yields relatively cautious prior predictions for the human toxicity. \\

Finally, we note our Bayesian decision-theoretic approach is applicable when animal data are collected from a number of preclinical studies in the same species.
In this case, preclinical information would be synthesised by meta-analysis to derive the informative component of the mixture prior and obtain optimal prior predictions for the human toxicity outcomes.
Discussion of using preclinical data collected from multiple animal species is beyond the scope of this paper. In this more complex setting, we may wish to allocate larger weights to species known to be more relevant to humans, whereas the decision-theoretic approach proposed at present does not allow us to incorporate this distinction.
This is one area for the future work. We are also exploring how we can use animal pharmacokinetic information to establish a Bayesian dose-exposure-toxicity model in light of the increasing interest in better understanding and characterisation of dose-toxicity relationship (FDA, 2003; Pinheiro and Duffull, 2009).

\newpage

\paragraph{Acknowledgements}

The authors would like to thank two anonymous reviewers and the Associate Editor for their helpful comments. 
This project has received funding from the European Union's Horizon 2020 research and innovation programme under the Marie Sk\l{}odowska-Curie grant agreement No 633567. Dr Hampson's contribution to this manuscript was supported by the UK Medical Research Council (grant MR/M013510/1).

\section*{Appendix}

\subsection*{A. Deriving the marginal probability density function for $p_j$}
\label{appd:mprobpj}

We consider to express the preclinical animal data for predicting the risks of toxicity at human doses as pseudo-data.
Thus, at these two pseudo dose levels $j=-1, 0$, uncertainty surrounding the risks of toxicity $p_j$ could be described with Beta distributions with parameters $t_{\mathcal{A} j}$ and $\nu_{\mathcal{A}j}$.
The joint prior probability density function (pdf) of $p_{-1}$ and $p_0$ is given by
$$
f(p_{-1}, p_0 \mid \boldsymbol{x}_\mathcal{A}) = \prod_{j=-1}^0 \frac{p_j^{t_{\mathcal{A} j}-1}(1-p_j)^{\nu_{\mathcal{A} j}-1}}{B(t_{\mathcal{A} j}, \nu_{\mathcal{A} j})},
$$
where $B(\cdot, \cdot)$ is the beta function. \\

\noindent Given the logistic dose-toxicity model, the joint pdf $f(p_{-1}, p_0)$ can be expressed in terms of the model parameters $\theta_1$ and $\theta_2$ via Jacobian transformation,
\begin{equation}
h(\theta_1, \theta_2 \mid \boldsymbol{x}_\mathcal{A})  = f(p_{-1}, p_0 \mid \boldsymbol{x}_\mathcal{A}) \times \frac{\partial (p_{-1}, p_0)}{\partial (\theta_1, \theta_2)}.
\end{equation}
\noindent From
$$
\log \left( \frac{p_j}{1-p_j}\right) = \theta_1 + \exp(\theta_2)\log(d_j/d_{\rm Ref}),
$$
we can easily derive
$$
\frac{\partial p_j}{\partial \theta_1} = p_j (1-p_j) \quad {\rm and} \quad \frac{\partial p_j}{\partial \theta_2} = p_j(1-p_j)\exp(\theta_2) \log(d_j /d_{\rm Ref}).
$$
Thus, the joint prior pdf of $\theta_1$ and $\theta_2$ can be written as
$$
h(\theta_1, \theta_2 \mid \boldsymbol{x}_\mathcal{A}) = \exp(\theta_2)\left| \log \left( \frac{d_{-1}}{d_0} \right) \right| \times \prod_{j=-1}^0 \frac{p_j^{t_{\mathcal{A} j}}(1-p_j)^{\nu_{\mathcal{A} j}}}{B(t_{\mathcal{A} j}, \nu_{\mathcal{A} j})},
$$
where the two pseudo doses $d_{-1}$ and $d_0$ correspond to the lowest and highest human doses in our context.
Substituting the $p_j$ with the logistic model parameters, we can write this joint prior pdf more explicitly:
\begin{equation*}
h(\theta_1, \theta_2 \mid \boldsymbol{x}_\mathcal{A}) = \exp(\theta_2) \left| \log \left( \frac{d_{-1}}{d_0} \right) \right|
\times \prod_{j=-1}^0 \frac{[1+\exp(-z_j)]^{-t_{\mathcal{A} j}}[1+\exp(z_j)]^{-\nu_{\mathcal{A} j}}}{B(t_{\mathcal{A} j}, \nu_{\mathcal{A} j})},
\end{equation*}
\noindent where $z_j = \theta_1 +\exp(\theta_2)\log(d_j/d_{\rm Ref})$. \\

\noindent By applying Jacobian transformation again, we can further derive the joint prior pdf of $p_i$ and $\theta_2$; for $i=1, \dots, I$,
\begin{equation*}
 g_i(p_i, \theta_2 \mid \boldsymbol{x}_\mathcal{A})  = h(\theta_1, \theta_2 \mid \boldsymbol{x}_\mathcal{A}) \times \frac{\partial (\theta_1, \theta_2)}{\partial (p_i, \theta_2)}.
\end{equation*}
\noindent With
\begin{equation*}
\left\{ \begin{aligned}
\theta_1 &= \log \left( \frac{p_i}{1-p_i}\right) - \exp(\theta_2) \log (d_i / d_{\rm Ref}) \\
\theta_2 &= \theta_2
\end{aligned} \right.
\end{equation*}
\noindent we can write
\begin{eqnarray*}
\frac{\partial \theta_1}{\partial p_i} = \frac{1}{p_i (1-p_i)}, &&
\frac{\partial \theta_1}{\partial \theta_2} = 0 , \\
\frac{\partial \theta_2}{\partial p_i} = 0, && \frac{\partial \theta_2}{\partial \theta_2} = 1,
\end{eqnarray*}
\noindent such that
\begin{equation*}
\begin{split}
 g_i(p_i, \theta_2 \mid \boldsymbol{x}_\mathcal{A})  &= h(\theta_1, \theta_2 \mid \boldsymbol{x}_\mathcal{A}) \times \frac{\partial (\theta_1, \theta_2)}{\partial (p_i, \theta_2)}, \\
 &= \frac{1}{p_i(1-p_i)} \cdot \exp(\theta_2)\left| \log \left( \frac{d_{-1}}{d_0} \right) \right| \times \prod_{j=-1}^0 \frac{[1+\exp(-z_{ji})]^{-t_{\mathcal{A} j}}[1+\exp(z_{ji})]^{-\nu_{\mathcal{A} j}}}{B(t_{\mathcal{A} j}, \nu_{\mathcal{A} j})},
\end{split}
\end{equation*}
\noindent where $z_{ji} = \theta_1 +\exp(\theta_2)\log(d_j/d_{\rm Ref})$.\\

\noindent Because $\theta_1$ in $z_j$ can be expressed with $\theta_2$ and $p_i$ given the logistic regression model that
$$
z_{ji} =  \log \left( \frac{p_i}{1-p_i}\right) + \exp(\theta_2) \log\left(\frac{d_j}{d_i}\right),
$$
the joint prior pdf $g_i(p_i, \theta_2)$ can therefore be parameterised with only $p_i$ and $\theta_2$.
The marginal probability density function for $p_j$, the risk of toxicity at dose $d_i$, $i=1,\dots, I$, can then be derived by integrating out the nuisance parameter $\theta_2$:
$$
f_i(p_i \mid \boldsymbol{x}_\mathcal{A}) = \int g_i(p_i, \theta_2 \mid \boldsymbol{x}_\mathcal{A}) {\rm d} \theta_2.
$$

\section*{B. Implied percentiles on the scale of $p_j$, given a bivariate normal prior for $\boldsymbol{\theta}$}
\label{appd:bivnorm2pj}

For $\log \left( \frac{p}{1-p} \right) = z$, the 95\% credible interval for $p$ is bounded by $\left( \frac{\exp(z_L)}{1+\exp(z_L)}, \frac{\exp(z_U)}{1+\exp(z_U)} \right)$ should we have known the lower and upper limits of $z$. Here $z$ can be seen as a transformed random variable, as following our parameterisation $z = \theta_1 + \exp(\theta_2) \log(d /d_\text{Ref})$.

\subsection*{B.1. The first two moments of the transformed random variable $z$}

The expectation for $z$ is $\mathbf{E}(z) = \mathbf{E}[\theta_1 + \exp(\theta_2) \log(d/d_\text{Ref})] = \mathbf{E}(\theta_1) + \log(d/d_\text{Ref}) \mathbf{E}(\exp(\theta_2))$.
By Taylor expansion, we know
\begin{equation*}
\begin{split}
\mathbf{E}(\exp(\theta_2)) &\approx \exp(\mathbf{E}(\theta_2)) + \frac{1}{2} \exp(\mathbf{E}(\theta_2))\cdot \text{Var}(\theta_2)  \\
&= \exp(\mathbf{E}(\theta_2))[1 + \frac{1}{2}\text{Var}(\theta_2)] \\
& \approx \exp\left[\mathbf{E}(\theta_2) + \frac{1}{2}\text{Var}(\theta_2) \right].
\end{split}
\end{equation*}
\noindent The last step follows the Taylor approximation $\exp(x) \approx 1+x$, which works well for small $x$. Having $x = \frac{1}{2}\text{Var}(\theta_2)$ leads to $\exp(\mathbf{E}(\theta_2))[1+x] \approx \exp(\mathbf{E}(\theta_2) + x)$.
Thus, the first moment for $z$ is approximated as
\begin{equation}
\label{eq:1stMoment}
\mathbf{E}(z) = \mathbf{E} (\theta_1) + \log(d/d_\text{Ref}) \exp\left[\mathbf{E}(\theta_2) + \frac{1}{2}\text{Var}(\theta_2) \right].
\end{equation}

Since $z^2 = \theta_1^2 + 2\theta_1\exp(\theta_2)\log(d/d_\text{Ref}) + \exp(2\theta_2)[\log(d/d_\text{Ref})]^2$, the second moment is then given by
\begin{equation}
\label{eq:2ndMoment}
\begin{split}
\mathbf{E}(z^2) & = \mathbf{E}(\theta_1^2) + 2\log(d/d_\text{Ref})\cdot \mathbf{E}(\theta_1 \cdot \exp(\theta_2)) + [\log(d/d_\text{Ref})]^2 \cdot \mathbf{E}(\exp(2\theta_2))   \\
& = \text{Var}(\theta_1) + [\mathbf{E}(\theta_1)]^2 + 2\log(d/d_\text{Ref})[\text{Cov}(\theta_1, \exp(\theta_2)) + \mathbf{E}(\theta_1)\mathbf{E}(\exp(\theta_2))]  \\
& \qquad \qquad \qquad + [\log(d/d_\text{Ref})]^2 \cdot \mathbf{E}(\exp(2\theta_2)),
\end{split}
\end{equation}

\noindent while we have
$$
[\mathbf{E}(z)]^2 = [\mathbf{E}(\theta_1)]^2 + 2\log(d/d_\text{Ref})\cdot \mathbf{E}(\theta_1)\mathbf{E}(\exp(\theta_2)) + [\log(d/d_\text{Ref})]^2 [\mathbf{E}(\exp(\theta_2))]^2.
$$
Thus,
\begin{equation}
\label{eq:varz}
\begin{split}
\text{Var}(z) &= \mathbf{E}(z^2) - [\mathbf{E}(z)]^2   \\
&= \text{Var}(\theta_1) + 2\log(d/d_\text{Ref})\cdot \text{Cov}(\theta_1, \exp(\theta_2)) \\
& \qquad  \qquad + [\log(d/d_\text{Ref})]^2 [\mathbf{E}(\exp(2\theta_2)) - \mathbf{E}(\exp(\theta_2))]^2 \\
& =\text{Var}(\theta_1) + 2\log(d/d_\text{Ref})\cdot \text{Cov}(\theta_1, \exp(\theta_2)) + [\log(d/d_\text{Ref})]^2 \cdot \text{Var}(\exp(\theta_2)).
\end{split}
\end{equation}

\noindent For the $\text{Cov}(\theta_1, \exp(\theta_2))$ in (\ref{eq:varz}), with Stein's Lemma, it holds that
\begin{equation*}
\begin{split}
\text{Cov}(\theta_1, \exp(\theta_2)) &= \mathbf{E}(\exp(\theta_2))\cdot \text{Cov}(\theta_1, \theta_2) \\
& \approx \exp\left[\mathbf{E}(\theta_2) + \frac{1}{2}\text{Var}(\theta_2) \right]\cdot \text{Cov}(\theta_1, \theta_2).
\end{split}
\end{equation*}

\noindent For the $\text{Var}(\exp(\theta_2))$ in (\ref{eq:varz}),
\begin{equation*}
\begin{split}
\text{Var}(\exp(\theta_2)) & = \mathbf{E}(\exp(2\theta_2)) - [\mathbf{E}(\exp(\theta_2))]^2 \\
& \approx \exp(2\mathbf{E}(\theta_2) + 2 \text{Var}(\theta_2)) - \exp(2\mathbf{E}(\theta_2) + \text{Var}(\theta_2) ) \\
& = \exp(2\mathbf{E}(\theta_2) + \text{Var}(\theta_2)) \cdot \exp(\text{Var}(\theta_2)) - \exp(2\mathbf{E}(\theta_2) + \text{Var}(\theta_2) ) \\
& = \exp(2\mathbf{E}(\theta_2) + \text{Var}(\theta_2)) [ \exp(\text{Var}(\theta_2)) - 1].
\end{split}
\end{equation*}

\subsection*{B.2. The lower and upper limits of $z$}

With (\ref{eq:1stMoment}) and (\ref{eq:varz}), the lower and upper limits of $z$ are
\begin{equation*}
\begin{split}
z_L &= \mathbf{E}(z) - 1.96\sqrt{\text{Var}(z)}, \\
z_U &= \mathbf{E}(z) + 1.96\sqrt{\text{Var}(z)}.
\end{split}
\end{equation*}

\noindent Obtaining the implied percentiles denoted by $q_{jk}'$, we can then easily code up the optimiser used to find a bivariate normal prior $\pi_0(\boldsymbol{\theta} \mid \boldsymbol{x}_\mathcal{A})$ given the prior probabilities $q_{jk}$ obtained following steps described in Section \ref{sec:BayesMod}.

\newpage

\section*{C. OpenBUGS code for implementation}
\label{appd:code}

\begin{verbatim}
model{
	# sampling model
	for(j in 1:Ncohorts){
		lin[j] <- theta[1] + exp(theta[2])*log(doseH[j]/dRef)
		logit(pTox[j]) <- lin[j]
		NtoxH[j] ~ dbin(pTox[j], NsubH[j])
	}
	
	for(i in 1:MdoseH){
		lin.star[i] <- theta[1] + exp(theta[2])*log(doseH[i]/dRef)
		logit(pTox.star[i]) <- lin.star[i]
		
		pCat[i, 1] <- step(pTox.cut[1] - pTox.star[i])
		pCat[i, 2] <- step(pTox.cut[2] - pTox.star[i])
		                  - step(pTox.cut[1] - pTox.star[i])
		pCat[i, 3] <- step(1 - pTox.star[i]) - step(pTox.cut[2] - pTox.star[i])	
	}


		theta[1:2] ~ dmnorm(thetaMu[which, 1:2], thetaPrec[which, 1:2, 1:2])
		which ~ dcat(wMix[1:2])
		# to monitor the exchangeability probability
		# in the course of the new human trial
			for(k in 1:2){
				prob.ex[k] <- equals(which, k)
			}

			
			thetaMu[1, 1:2] ~ dmnorm(PriorA[1:2], thetaPrec[1, 1:2, 1:2])
			cov.A[1, 1] <- PriorA[3]
			cov.A[1, 2] <- PriorA[4]
			cov.A[2, 1] <- cov.A[1, 2]
			cov.A[2, 2] <- PriorA[5]
			thetaPrec[1, 1:2, 1:2] <- inverse(cov.A[1:2, 1:2])


			thetaMu[2, 1:2] ~ dmnorm(Prior.mw[1:2], thetaPrec[2, 1:2, 1:2])
			cov.rb[1, 1] <- pow(Prior.sw[1], 2)
			cov.rb[2, 2] <- pow(Prior.sw[2], 2)
			cov.rb[1, 2] <- Prior.sw[1]*Prior.sw[2]*Prior.corr
			cov.rb[2, 1] <- cov.rb[1, 2]
			thetaPrec[2, 1:2, 1:2] <- inverse(cov.rb[1:2, 1:2])
}
\end{verbatim}


\newpage

\tikzstyle{block} = [draw, fill=gray!20, rectangle, 
    minimum height=6em, minimum width=4em]
\tikzstyle{sum} = [draw, fill=gray!20, circle, node distance=1cm]
\tikzstyle{input} = [coordinate]
\tikzstyle{output} = [coordinate]
\tikzstyle{pinstyle} = [pin edge={to-,thin,black}, minimum width=2em]

\begin{figure}
\centering
\begin{tikzpicture}[auto, >=latex']
    \node [block, align=center, text width=2.5cm, pin={above:For analysing an ongoing phase I trial}] (cohort) {Observe data from cohort $h=1,\dots,H$};
    \node [block, right of=cohort,
            node distance=5cm, text width=4.3cm] (compare) {Compare the observed human data from cohorts $1,\dots,h$, with prior predictions based on animal data};
     \node [block, right of=compare,
            node distance=6cm, text width=4cm] (mixture) {Determine the dynamic prior weight $w^{(h)}$ in the mixture prior $\mu_0^{(h)}(\boldsymbol{\theta} | \boldsymbol{x}_\mathcal{A})$};
     \node [block, below of=mixture,
            node distance=4cm, text width=4cm] (posterior) {Update  $\mu_0^{(h)}(\boldsymbol{\theta} | \boldsymbol{x}_\mathcal{A})$ with human data from cohorts $1,\dots,h$ to the mixture posterior $\mu_0^{(h)}(\boldsymbol{\theta} | \boldsymbol{x}_\mathcal{A}, \boldsymbol{x}_\mathcal{H}^{(h)})$};
     \node [block, below of=compare,
            node distance=4cm, text width=3.5cm] (recommend) {Recommend a dose for cohort $(h+1)$ based on the mixture posterior};
     \draw [->] (cohort) -- (compare);   
     \draw [->] (compare) -- (mixture);       
     \draw [->] (mixture) -- (posterior);    
     \draw [->] (posterior) -- (recommend);    
     \draw [->] (recommend) -| (cohort);  
\end{tikzpicture}
\caption{A schematic diagram of the Bayesian dose-escalation procedure driven by a dynamic mixture prior, where the prior mixture weight $w^{(h)}$ for using animal data is chosen based on the proposed decision-theoretic approach.}
\label{fig:schem}
\end{figure}
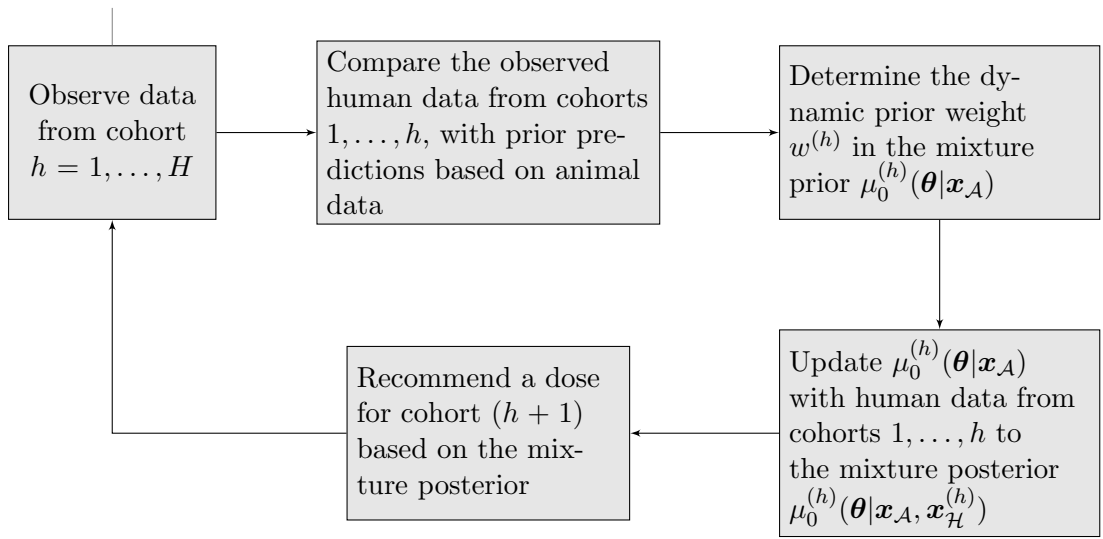

\begin{figure}
\centering
\captionsetup{font=scriptsize}
\includegraphics[width=1\linewidth]{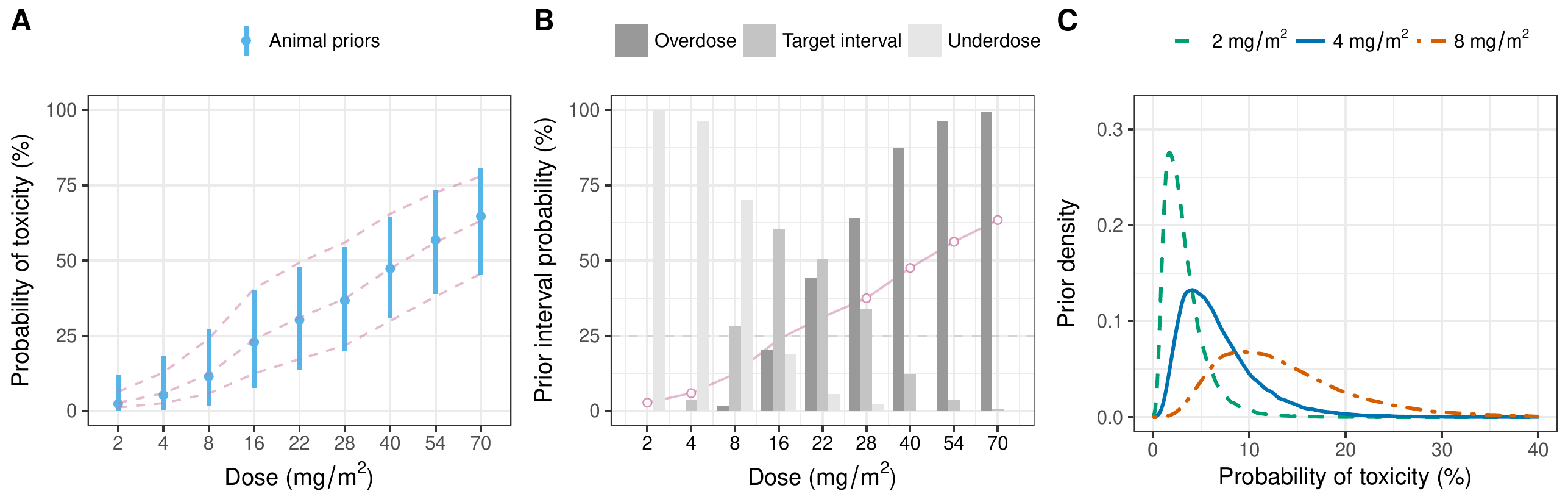}
\caption{Summaries of priors based on preclinical information. Panel A shows median and 95\% CI of the marginal prior distributions for $p_i$ in blue bars, together with the fitted probabilities in pink dashed lines from the bivariate normal prior $\pi_0(\boldsymbol{\theta} \mid \boldsymbol{x}_\mathcal{A})$ found by an optimiser. Panel B gives an overview of the interval probabilities, where the background red curve indicates the prior medians for probability of toxicity per dose. Panel C presents prior densities for $p_i$ at candidate starting doses.}
\label{fig:aPriors}
\end{figure}

\begin{figure}
\centering
\captionsetup{font=scriptsize}
\includegraphics[width=0.73\linewidth]{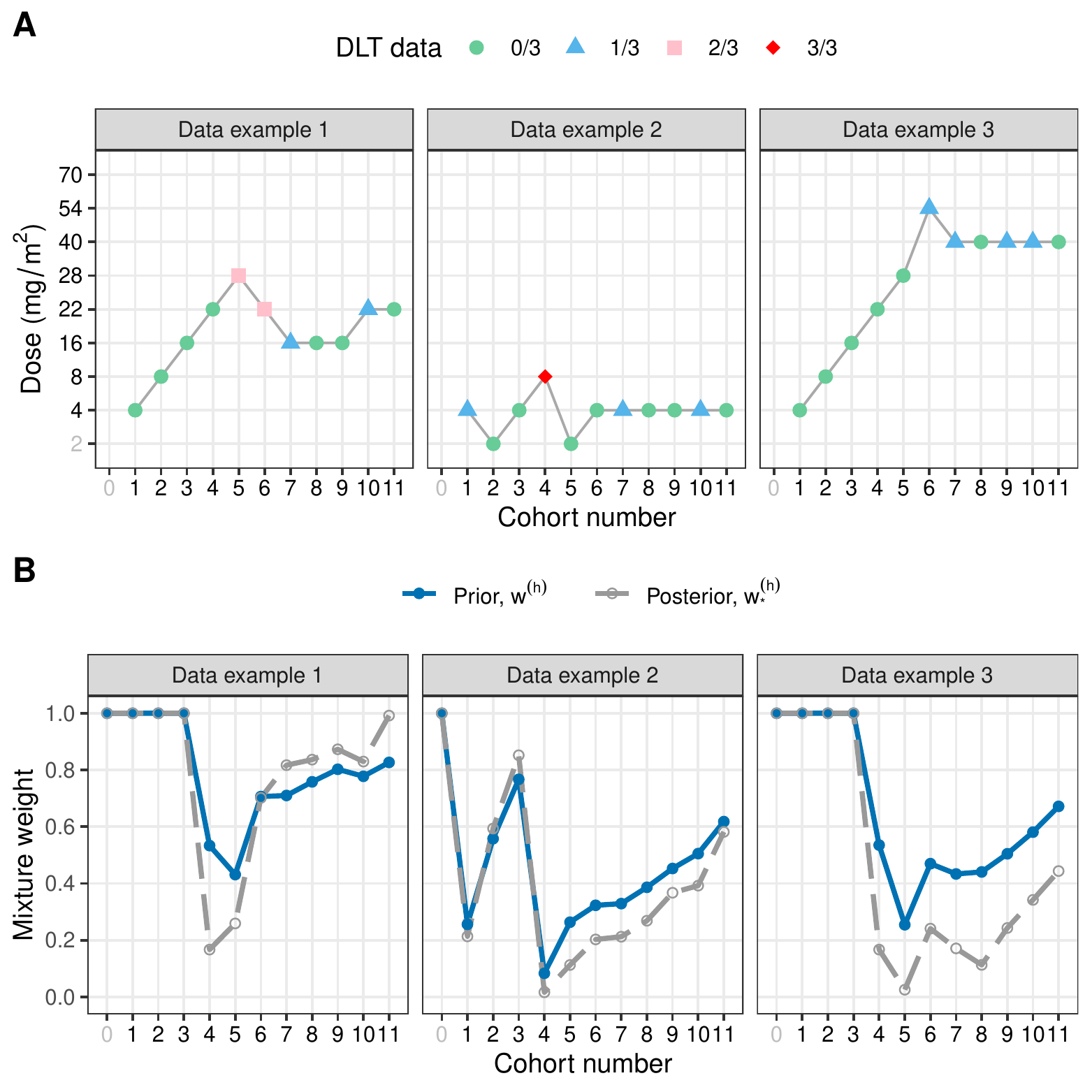}
\caption{Trajectory of dose recommendations (Panel A) and dynamic update of mixture weight attributed to preclinical information (Panel B) during the course of each hypothetical phase I clinical trial. Cohort number labelled as 0 refers to when the phase I trial has not yet started.}
\label{fig:dataSc}
\end{figure}

\begin{figure}
\centering
\captionsetup{font=scriptsize}
\includegraphics[width=0.73\linewidth]{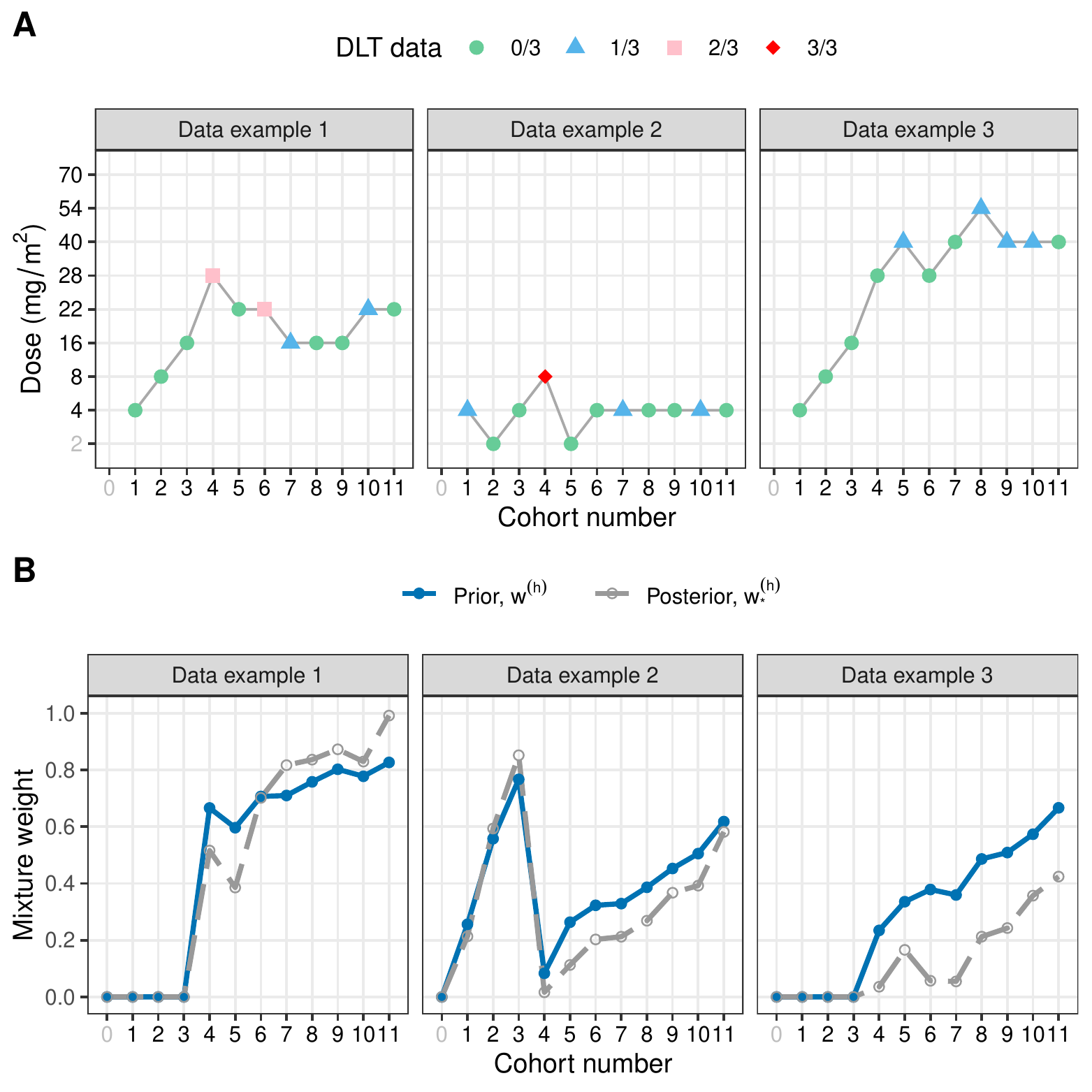}
\caption{Trajectory of dose recommendations (Panel A) and dynamic update of mixture weight attributed to preclinical information (Panel B) during the course of each hypothetical phase I clinical trial under a two-stage design, with a run-in period characterised in the first stage and dose-escalation procedure driven by a mixture prior in the second stage. Cohort number labelled as 0 refers to when the phase I trial has not yet started.}
\label{fig:dataScRunIn}
\end{figure}

\begin{figure}
\centering
\captionsetup{font=scriptsize}
\includegraphics[width=0.9\linewidth]{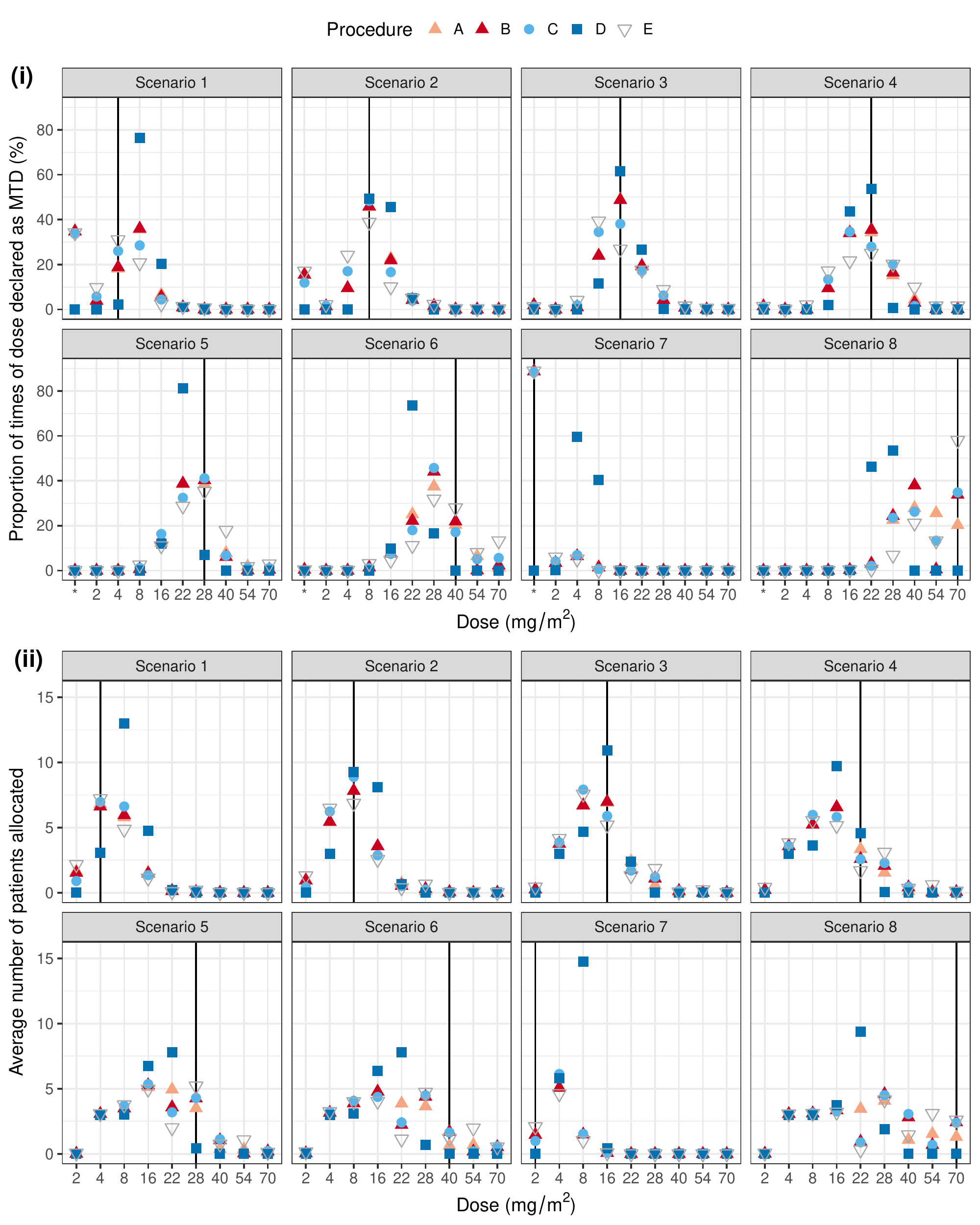}
\caption{Operating characteristics of phase I dose-escalation trials conducted using Procedures A - E. The vertical black line indicates the true MTD in humans in each simulation scenario. $^{\star}$ indicates the trial was stopped early for safety. Proportions and averages are calculated by averaging across the 1000 simulated trials in each scenario.}
\label{fig:PhI}
\end{figure}

\clearpage

\begin{table}
\scriptsize
\centering
\captionsetup{font=scriptsize}
\caption{Cross-tabulation of utilities for the predicted versus the observed human binary DLT outcomes.}
\begin{tabular}{@{\extracolsep{10pt}}clcc@{}}
\toprule
 &    & \multicolumn{2}{c}{Observation ($y$)}	 \\
\cline{3-4}  \\[-0.4em]
	&	&No-DLT 	&DLT   \\	
\hline \\[-0.8em]
Prior prediction ($\eta$) 	&No-DLT	&$u_{00}$		&$u_{10}$  	 	\\
	&DLT		&$u_{01}$ 	&$u_{11}$  			 \\
\bottomrule
\end{tabular}
\label{tab:2by2rp}
\end{table}

\begin{table}
\scriptsize
\centering
\captionsetup{font=scriptsize}
\caption{Effective sample sizes of the marginal prior distributions for DLT risks based on animal data.}
\begin{tabular}{@{}lccccccccc@{}}
\toprule
&\multicolumn{9}{@{}c}{Dose (mg/m$^2$)}  \\
\cline{2-10} \\[-0.4em]
&$d_{1}$  &$d_{2}$  &$d_{3}$  &$d_{4}$  &$d_{5}$  &$d_{6}$  &$d_{7}$ &$d_{8}$   &$d_{9}$    \\
& 2            & 4           & 8            & 16            & 22            & 28           & 40  		& 54			   & 70     \\
\midrule
Prior means    & 0.033   & 0.069   & 0.137  & 0.252   & 0.322   & 0.382    & 0.476  & 0.557    & 0.625 \\
Prior std dev.  & 0.022  & 0.041   & 0.070   & 0.102  & 0.115        & 0.121     & 0.125    & 0.121     & 0.114  \\
\cmidrule{1-10}
ESS          & 63.4   & 36.7 & 23.3 & 17.0 & 15.6 &  15.0 & 15.0 &  15.8 & 17.0   \\
\quad $a$    & 2.1  & 2.5  & 3.2  & 4.3  & 5.0  & 5.7  & 7.1  & 8.8  & 10.6    \\
\quad $b$    & 61.3 & 34.2 & 20.1 &  12.7 & 10.6 & 9.3  & 7.9  &  7.0  & 6.4   \\
\bottomrule
\end{tabular}
\label{tab:pseudopts}
\end{table}

\begin{table}
\scriptsize
\centering
\captionsetup{font=scriptsize}
\caption{Simulation scenarios for the true DLT risks in humans. The figure in bold indicates the true MTD.}
\begin{tabular}{@{\extracolsep{12pt}}lccccccccc@{}}
\toprule
&\multicolumn{9}{@{}c}{Dose (mg/m$^2$)}  \\
\cline{2-10} \\[-0.4em]
&$d_{1}$  &$d_{2}$  &$d_{3}$  &$d_{4}$  &$d_{5}$  &$d_{6}$  &$d_{7}$ &$d_{8}$   &$d_{9}$   \\
& 2            & 4           & 8            & 16            & 22            & 28           & 40  		& 54			   & 70    \\
\midrule
Scenario 1  	 & 0.11   & \textbf{0.25}   & 0.35   & 0.41   & 0.47   & 0.52   & 0.58   & 0.63   & 0.70 \\
Scenario 2     & 0.08   & 0.16   & \textbf{0.25}   & 0.35   & 0.42   & 0.45   & 0.53   & 0.60   & 0.70  \\
Scenario 3     & 0.02   & 0.05   & 0.14   & \textbf{0.25}   & 0.35   & 0.42   & 0.51   & 0.60   & 0.68  \\
Scenario 4     & 0.03   & 0.05   & 0.10   & 0.16   & \textbf{0.25}   & 0.32   & 0.40   & 0.48   & 0.55  \\
Scenario 5     & 0.001  & 0.005  & 0.03   & 0.10   & 0.16   & \textbf{0.25}   & 0.38   & 0.50   & 0.60  \\
Scenario 6     & 0.01   & 0.02   & 0.05   & 0.08   & 0.11   & 0.14   & \textbf{0.25}   & 0.37   & 0.47  \\
Scenario 7     & \textbf{0.35}   & 0.42   & 0.60   & 0.75   & 0.82   & 0.88   & 0.91   & 0.94   & 0.97  \\
Scenario 8     & 0.001  & 0.005  & 0.01   & 0.02   & 0.04   & 0.05   & 0.10   & 0.16   & \textbf{0.25}  \\
\bottomrule
\end{tabular}
\label{tab:base}
\end{table}

\clearpage

\bibliographystyle{apalike}





\section*{Supplementary material}

# Web-based Supplementary Materials for:
A Bayesian decision-theoretic approach to incorporate preclinical information into phase I oncology trials

**by Haiyan Zheng, Lisa V. Hampson**

**A. DATA EXAMPLES FOR NO BORROWING OR FULL POOLING OF ANIMAL DATA**

In Section 5.2, we simulated three hypothetical phase I clinical trials to exemplify interim dose recommendations, using the proposed method to leverage preclinical data without undermining patients' safety. Here, we show in Figure S1 how doses would have been recommended in an alternative Bayesian dose-escalation procedure driven by either an operational prior $m_0(\boldsymbol{\theta})$ or an animal prior $\pi_0(\boldsymbol{\theta}|\boldsymbol{x}_{\mathcal{A}})$.

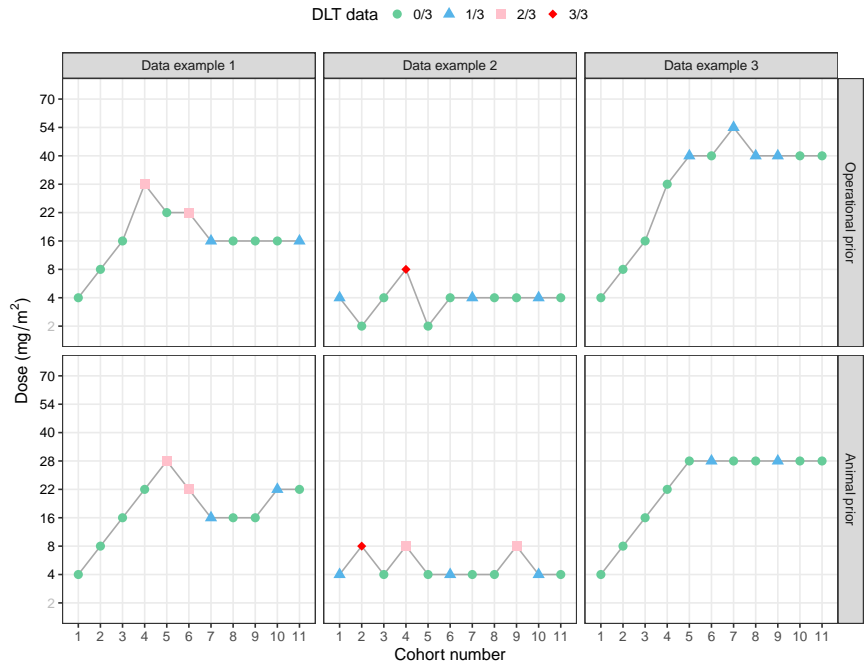

Figure S1: Trajectory of dose recommendations under alternative Bayesian dose-escalation procedures.

These new data examples presented in Figure S1 were simulated from the same parameter settings used for those presented in Figure 3 of the main manuscript for a fair comparison: a vector of binary outcomes on each dose were simulated and then sampled without replacement as each new patient was assigned to a dose in the dose-escalation study. For example, looking at data examples with the same label, the first patient receiving dose 28 mg/m² in Figure 3 and Figure S1, involved in different Bayesian dose-escalation procedures will have the same simulated binary toxicity outcome. We may therefore observe what could have been the consequence by adopting different priors in a Bayesian dose-escalation procedure.

From this comparison, we can see the impact of leveraging the preclinical data on decision making in an adaptive phase I clinical trial. Using our approach leads to compromises between

1

full pooling and complete discard of preclinical animal data. In the simulated trials labelled with data example 1, we observe that behaviours of the trial using our approach is similar with that implememted with Bayesian approach fully incorporating animal data in the prior, as under this prior-data consistency scenario a large prior mixture weight will be attributed to animal data based upon our assessment of commensurability. Advantages of using our approach are also evident in scenarios of a prior-data conflict: unlike a trial with animal data fully incorporated, less patients were allocated with overly toxic dose 8 mg/m$^2$ in data example 2, while more patients will have chance to escalate to a true target dose which is dose 40 mg/m$^2$ in data example 3.

## B. PRIOR MIXTURE WEIGHT SPECIFIC TO COHORT $h$

In Sections 5.2 and 5.3, we presented the trajectories of dose recommendations together with the dynamic changes of prior mixture weight $w^{(h)}$ attributed to the informative component at each stage of a hypothetical phase I trial. Here, we list the computed commensurability $\kappa^{(h)}$ across the doses that have been given to patients thus far, as well as the tuning parameter $\lambda^{(h)}$, in below the Table S1 and S2, for the unconstrained and constrained versions of the proposed Bayesian decision-theoretic approach, respectively. The prior mixture weight $w^{(h)}$ visualised on Figures 4B) and 5B) of the main paper is a function of these two arguments.

Table S1: The assessed commensurability $\kappa^{(h)}$ and the tuning parameter $\lambda^{(h)}$ are updated as the trial progresses, if implementing the unconstrained version (without a run-in period) of the proposed approach to compute the prior mixture weight $w^{(h)}$.

|  |  | Cohort number, $h$ | | | | | | | | | | |
| --- | --- | --- | --- | --- | --- | --- | --- | --- | --- | --- | --- | --- |
|  |  | 1 | 2 | 3 | 4 | 5 | 6 | 7 | 8 | 9 | 10 | 11 |
| Data example 1 | $\kappa^{(h)}$ | 1.00 | 1.00 | 1.00 | 0.80 | 0.73 | 0.87 | 0.86 | 0.87 | 0.88 | 0.84 | 0.83 |
|  | $\lambda^{(h)}$ | 3.36 | 3.21 | 3.04 | 2.82 | 2.71 | 2.43 | 2.24 | 2.02 | 1.71 | 1.43 | 1.01 |
|  | $w^{(h)}$ | 1.00 | 1.00 | 1.00 | 0.53 | 0.43 | 0.71 | 0.71 | 0.76 | 0.80 | 0.78 | 0.83 |
| Data example 2 | $\kappa^{(h)}$ | 0.67 | 0.83 | 0.92 | 0.42 | 0.61 | 0.63 | 0.61 | 0.62 | 0.63 | 0.62 | 0.62 |
|  | $\lambda^{(h)}$ | 3.36 | 3.20 | 3.05 | 2.84 | 2.71 | 2.44 | 2.26 | 2.00 | 1.71 | 1.42 | 1.00 |
|  | $w^{(h)}$ | 0.26 | 0.56 | 0.77 | 0.08 | 0.26 | 0.32 | 0.33 | 0.39 | 0.45 | 0.50 | 0.62 |
| Data example 3 | $\kappa^{(h)}$ | 1.00 | 1.00 | 1.00 | 0.80 | 0.60 | 0.73 | 0.69 | 0.67 | 0.67 | 0.68 | 0.67 |
|  | $\lambda^{(h)}$ | 3.38 | 3.22 | 3.01 | 2.8 | 2.68 | 2.44 | 2.24 | 2.02 | 1.73 | 1.40 | 1.00 |
|  | $w^{(h)}$ | 1.00 | 1.00 | 1.00 | 0.54 | 0.25 | 0.47 | 0.43 | 0.44 | 0.50 | 0.58 | 0.67 |

Table S2: The assessed commensurability $\kappa^{(h)}$ and the tuning parameter $\lambda^{(h)}$ are updated as the trial progresses, if implementing the constrained version (with a run-in period) of the proposed approach to compute the prior mixture weight $w^{(h)}$.

|  |  | Cohort number, $h$ | | | | | | | | | | |
| --- | --- | --- | --- | --- | --- | --- | --- | --- | --- | --- | --- | --- |
|  |  | 1 | 2 | 3 | 4 | 5 | 6 | 7 | 8 | 9 | 10 | 11 |
| Data example 1 | $\kappa^{(h)}$ | 1.00 | 1.00 | 1.00 | 0.87 | 0.82 | 0.87 | 0.86 | 0.87 | 0.88 | 0.84 | 0.83 |
|  | $\lambda^{(h)}$ | 0 | 0 | 0 | 2.84 | 2.64 | 2.43 | 2.24 | 2.02 | 1.71 | 1.43 | 1.01 |
|  | $w^{(h)}$ | 0 | 0 | 0 | 0.67 | 0.60 | 0.71 | 0.71 | 0.76 | 0.80 | 0.78 | 0.83 |
| Data example 2 | $\kappa^{(h)}$ | 0.67 | 0.83 | 0.92 | 0.42 | 0.61 | 0.63 | 0.61 | 0.62 | 0.63 | 0.62 | 0.62 |
|  | $\lambda^{(h)}$ | 3.36 | 3.20 | 3.05 | 2.84 | 2.71 | 2.44 | 2.26 | 2.00 | 1.71 | 1.42 | 1.00 |
|  | $w^{(h)}$ | 0.26 | 0.56 | 0.77 | 0.08 | 0.26 | 0.32 | 0.33 | 0.39 | 0.45 | 0.50 | 0.62 |
| Data example 3 | $\kappa^{(h)}$ | 1.00 | 1.00 | 1.00 | 0.60 | 0.67 | 0.67 | 0.63 | 0.70 | 0.67 | 0.68 | 0.67 |
|  | $\lambda^{(h)}$ | 0 | 0 | 0 | 2.84 | 2.69 | 2.39 | 2.24 | 2.02 | 1.71 | 1.43 | 1.02 |
|  | $w^{(h)}$ | 0 | 0 | 0 | 0.24 | 0.34 | 0.38 | 0.36 | 0.49 | 0.51 | 0.57 | 0.67 |

## C. NUMERICAL RESULTS OF ALL EVALUATE SCENARIOS

The performance of trials using BLRM-guided dose-escalation Procedures A – E are compared with that of the optimal non-parametric benchmark design by [1]. The optimal design is defined using the 'complete' toxicity profile of each patient, created by assuming there are $J_{i^\star}$ clones of a patient given doses spanning the dosing set $\mathcal{D}_{i^\star}$. A toxicity tolerance thereshold $\epsilon_n$ is generated from $U[0,1]$ for the $n$th patient, which determines the corresponding toxicity outcome at the $j$th dose as

$$R_{jn} = \mathbb{1}(\epsilon_n \leq p_{i^\star j}), \quad 1 \leq n \leq N, \quad 1 \leq j \leq J_{i^\star},$$

where $\mathbb{1}(\cdot)$ is the indicator function. An unbiased estimate for $p_{i^\star j}$ is thus $\bar{R}_j(N) = \frac{1}{N}\sum_{n=1}^{N} R_{jn}$ for a trial of which the maximum sample size is $N$. Consequently, the estimated MTD under the benchmark design is

$$\hat{d}_M^{\text{opt}} = \arg\min_{j=1,\ldots,J_{i^\star}} |\bar{R}_j(N) - 0.25|.$$

Improvements beyond this bound are not possible unless strong parametric assumptions are made about dose-response relationships. In our context, we wish to quantify the gains that can be made over the benchmark designs, in part due to borrowing strength from the preclinical data. Table S3 provides a complete listing of all simulation results for analysis models defined in Section 6 and the optimal benchmark design.

## D. ADDITIONAL SIMULATION RESULTS

We visualise additional results of the simulation study in Figure S2. In what is presented here, we have set $\lambda^{(h)} = \sqrt{N/n^{(h)}}$ together with $u_{01} = 0.6$ to compute the prior mixture weight $w^{(h)}$ attributed to the informative preclinical component in the mixture prior. Each simulated trial has maximum 21 patients; that is, in total seven cohorts. Results show that Procedures A and B stipulating the tuning parameter $\lambda^{(h)} = \sqrt{N/n^{(h)}}$ work equally well as their variants stipulating $\lambda^{(h)}$ in the form of (14) defined in the main manuscript. Figure S3 shows the results from another simulation study, where we have set the maximum trial sample size as 33 (i.e., 11 patient cohorts). With increase of trial sample size, differences between Procedures A and B for the PCS in scenario 8. As was explained in the main manuscript, this is because we will have chance to escalate to the highest dose when more information accrues from the current trial to overcome the incompatible animal prior information.

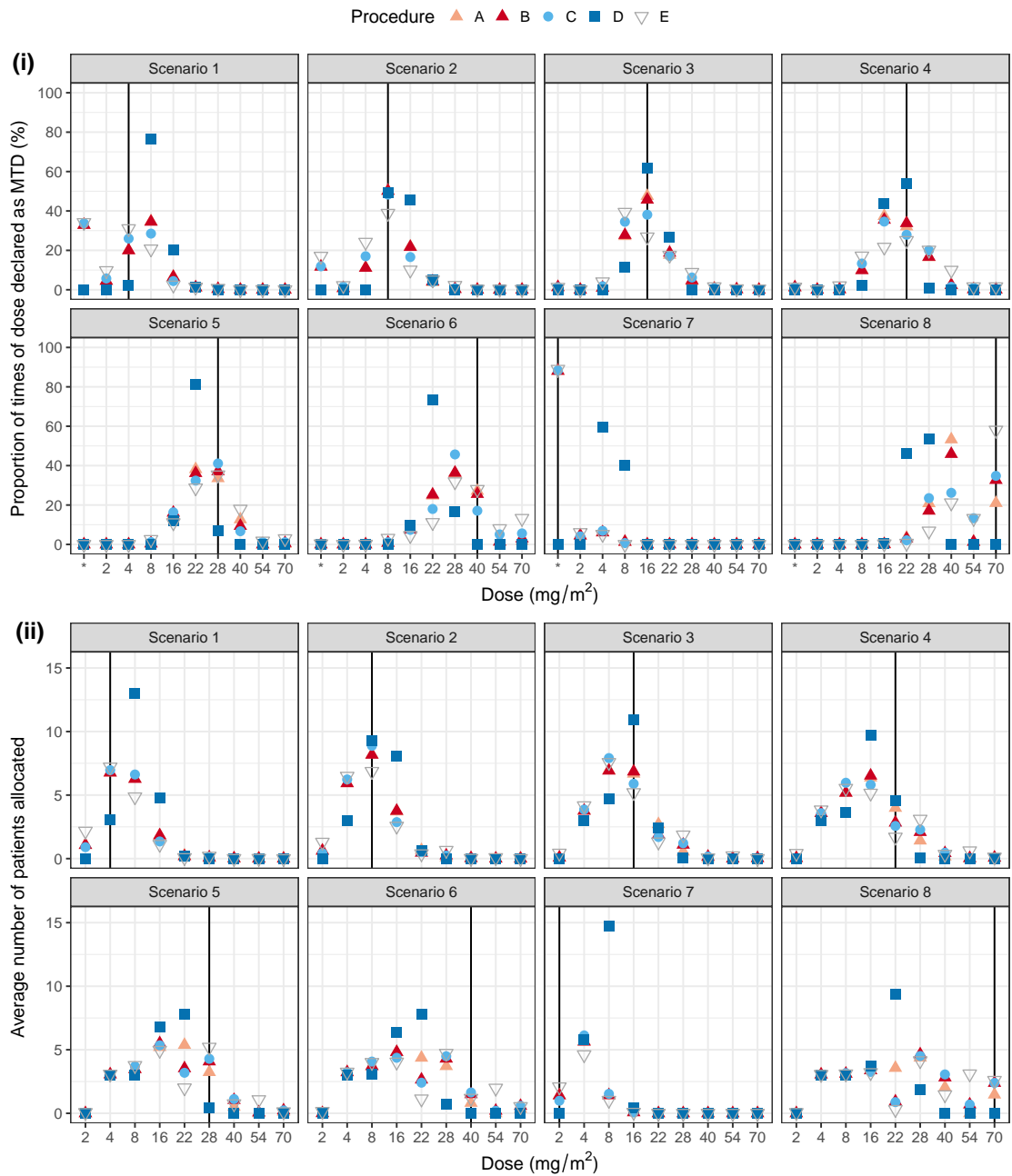

Figure S2: Operating characteristics of phase I dose-escalation trials conducted using Procedures A - E, where the tuning parameter is stipulated as $\lambda^{(h)} = \sqrt{N/n^{(h)}}$ for Procedures A and B. The vertical black line indicates the true MTD in humans in each simulation scenario.

4

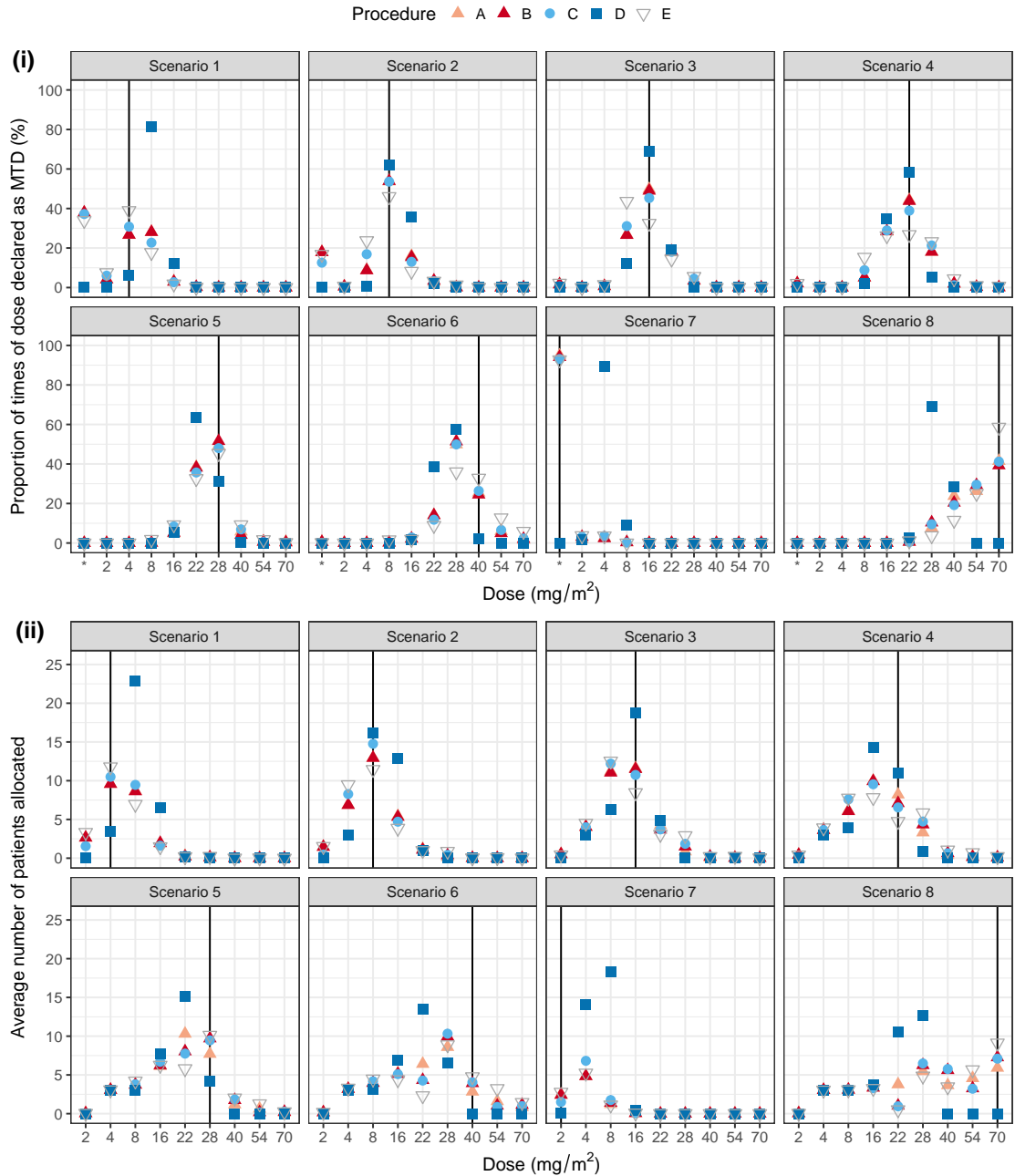

Figure S3: Operating characteristics of phase I dose-escalation trials conducted using Procedures A - E, when the maximum trial sample size is 33 (i.e., 11 patient cohorts). The vertical black line indicates the true MTD in humans in each simulation scenario.

5

Table S3: Comparison of the analysis models in terms of the percentage of selecting a dose as MTD at the end of the trials, percentage of early stopping for safety, average patient allocation, and average number of patients with toxicity.

| Sc. | Design | | | % dose declared as MTD & average patient allocation | | | | | | | | | | DLT | $\bar{N}$ |
|---|---|---|---|---|---|---|---|---|---|---|---|---|---|---|---|
| | | | | $d_{i^\star 1}$ 2 | $d_{i^\star 2}$ 4 | $d_{i^\star 3}$ 8 | $d_{i^\star 4}$ 16 | $d_{i^\star 5}$ 22 | $d_{i^\star 6}$ 28 | $d_{i^\star 7}$ 40 | $d_{i^\star 8}$ 54 | $d_{i^\star 9}$ 70 | None | | |
| 1 | | | pTox | 0.11 | **0.25** | 0.35 | 0.41 | 0.47 | 0.52 | 0.58 | 0.63 | 0.70 | | | |
| | Optimal | | Sel | 18.2 | **54.2** | 19.8 | 5.6 | 1.5 | 0.4 | 0.1 | 0 | 0 | | | |
| | A | $u_{01} = 0.6$ | Sel | 4.2 | **18.2** | 35.7 | 6.6 | 0.7 | 0.1 | 0 | 0 | 0 | 34.5 | | |
| | | | Pts | 1.6 | 6.8 | 5.8 | 1.6 | 0.2 | 0 | 0 | 0 | 0 | | 4.7 | 16.0 |
| | | $u_{01} = 0.2$ | Sel | 4.2 | **18.2** | 35.8 | 6.4 | 0.8 | 0.1 | 0 | 0 | 0 | 34.5 | | |
| | | | Pts | 1.6 | 6.8 | 5.8 | 1.5 | 0.2 | 0 | 0 | 0 | 0 | | 4.7 | 15.9 |
| | B | $u_{01} = 0.6$ | Sel | 3.8 | **18.8** | 36.0 | 5.6 | 1.0 | 0 | 0 | 0 | 0 | 34.8 | | |
| | | | Pts | 1.5 | 6.6 | 6.0 | 1.5 | 0.1 | 0.1 | 0 | 0 | 0 | | 4.6 | 15.8 |
| | | $u_{01} = 0.2$ | Sel | 3.8 | **18.8** | 35.9 | 5.7 | 1.0 | 0 | 0 | 0 | 0 | 34.8 | | |
| | | | Pts | 1.5 | 6.6 | 6.0 | 1.4 | 0.2 | 0.1 | 0 | 0 | 0 | | 4.7 | 15.8 |
| | C | | Sel | 5.7 | **26.0** | 28.5 | 4.4 | 1.3 | 0.4 | 0 | 0 | 0 | 33.7 | | |
| | | | Pts | 0.9 | 7.0 | 6.6 | 1.4 | 0.1 | 0.1 | 0 | 0 | 0 | | 4.8 | 16.1 |
| | D | | Sel | 0 | **2.2** | 76.4 | 20.3 | 1.1 | 0 | 0 | 0 | 0 | 0 | | |
| | | | Pts | 0 | 3.1 | 13.0 | 4.7 | 0.2 | 0 | 0 | 0 | 0 | | 7.3 | 21.0 |
| | E | | Sel | 9.7 | **31.0** | 20.7 | 2.4 | 1.3 | 0.7 | 0 | 0.1 | 0 | 34.1 | | |
| | | | Pts | 2.2 | 7.2 | 4.9 | 1.1 | 0.1 | 0.2 | 0 | 0 | 0 | | 4.3 | 15.7 |
| 2 | | | pTox | 0.08 | 0.16 | **0.25** | 0.35 | 0.41 | 0.45 | 0.52 | 0.58 | 0.63 | | | |
| | Optimal | | Sel | 4.3 | 28.1 | **39.8** | 20.3 | 4.9 | 1.7 | 0.7 | 0.1 | 0.1 | | | |
| | A | $u_{01} = 0.6$ | Sel | 1.4 | 9.5 | **45.6** | 22.7 | 4.4 | 0.8 | 0.1 | 0 | 0 | 15.5 | | |
| | | | Pts | 1.0 | 5.4 | 7.8 | 3.6 | 0.7 | 0.1 | 0 | 0 | 0 | | 4.5 | 18.6 |
| | | $u_{01} = 0.2$ | Sel | 1.4 | 9.5 | **46.3** | 21.8 | 4.6 | 0.9 | 0 | 0 | 0 | 15.5 | | |
| | | | Pts | 1.0 | 5.4 | 7.8 | 3.6 | 0.7 | 0.1 | 0 | 0 | 0 | | 4.5 | 18.6 |
| | B | $u_{01} = 0.6$ | Sel | 1.4 | 9.5 | **45.9** | 21.9 | 4.2 | 1.6 | 0 | 0 | 0 | 15.5 | | |
| | | | Pts | 1.0 | 5.4 | 7.8 | 3.6 | 0.5 | 0.3 | 0.1 | 0 | 0 | | 4.5 | 18.7 |
| | | $u_{01} = 0.2$ | Sel | 1.4 | 9.5 | **46.0** | 21.4 | 5.2 | 0.9 | 0.1 | 0 | 0 | 15.5 | | |
| | | | Pts | 1.0 | 5.4 | 7.8 | 3.6 | 0.6 | 0.3 | 0.1 | 0 | 0 | | 4.5 | 18.8 |
| | C | | Sel | 1.5 | 17.0 | **48.3** | 16.6 | 3.9 | 0.7 | 0.1 | 0 | 0 | 11.9 | | |
| | | | Pts | 0.4 | 6.2 | 8.9 | 2.9 | 0.5 | 0.2 | 0 | 0 | 0 | | 4.6 | 19.1 |
| | D | | Sel | 0 | 0 | **49.3** | 45.7 | 5.0 | 0 | 0 | 0 | 0 | 0 | | |
| | | | Pts | 0 | 3.0 | 9.3 | 8.1 | 0.6 | 0 | 0 | 0 | 0 | | 5.8 | 21.0 |
| | E | | Sel | 2.1 | 24.1 | **38.8** | 10.0 | 5.0 | 2.2 | 0.3 | 0.3 | 0.2 | 17.0 | | |
| | | | Pts | 1.3 | 6.5 | 6.9 | 2.5 | 0.4 | 0.6 | 0 | 0.1 | 0 | | 4.2 | 18.3 |
| 3 | | | pTox | 0.02 | 0.05 | 0.14 | **0.25** | 0.35 | 0.42 | 0.51 | 0.60 | 0.68 | | | |
| | Optimal | | Sel | 0 | 1.0 | 24.5 | **46.3** | 21.5 | 5.3 | 1.3 | 0.1 | 0 | | | |
| | A | $u_{01} = 0.6$ | Sel | 0 | 1.0 | 24.0 | **49.0** | 19.0 | 4.1 | 0.9 | 0.1 | 0 | 1.9 | | |
| | | | Pts | 0.2 | 3.8 | 6.7 | 6.9 | 2.4 | 0.6 | 0 | 0 | 0 | | 3.9 | 20.6 |
| | | $u_{01} = 0.2$ | Sel | 0 | 1.0 | 24.9 | **47.1** | 21.2 | 3.9 | 0 | 0 | 0 | 1.9 | | |
| | | | Pts | 0.2 | 3.8 | 6.8 | 7.0 | 2.5 | 0.4 | 0 | 0 | 0 | | 4.5 | 20.7 |
| | B | $u_{01} = 0.6$ | Sel | 0 | 1.0 | 23.9 | **48.7** | 19.4 | 4.3 | 0.8 | 0 | 0 | 1.9 | | |
| | | | Pts | 0.2 | 3.8 | 6.7 | 7.0 | 1.8 | 1.1 | 0.1 | 0 | 0 | | 4.0 | 20.7 |
| | | $u_{01} = 0.2$ | Sel | 0 | 1.0 | 24.6 | **47.6** | 21.1 | 3.7 | 0.1 | 0 | 0 | 1.9 | | |
| | | | Pts | 0.2 | 3.8 | 6.7 | 7.0 | 1.8 | 1.1 | 0.1 | 0 | 0 | | 4.0 | 20.7 |
| | C | | Sel | 0 | 1.8 | 34.5 | **38.1** | 17.1 | 6.2 | 0.7 | 0.2 | 0.1 | 1.3 | | |
| | | | Pts | 0 | 3.9 | 7.9 | 5.9 | 1.7 | 1.2 | 0.2 | 0 | 0 | | 4.0 | 20.8 |
| | D | | Sel | 0 | 0 | 11.6 | **61.6** | 26.7 | 0.1 | 0 | 0 | 0 | 0 | | |
| | | | Pts | 0 | 3.0 | 4.7 | 10.9 | 2.4 | 0 | 0 | 0 | 0 | | 4.3 | 21.0 |

6

| Sc. | Design | | | $d_{i^\star 1}$ 2 | $d_{i^\star 2}$ 4 | $d_{i^\star 3}$ 8 | $d_{i^\star 4}$ 16 | $d_{i^\star 5}$ 22 | $d_{i^\star 6}$ 28 | $d_{i^\star 7}$ 40 | $d_{i^\star 8}$ 54 | $d_{i^\star 9}$ 70 | None | DLT | $\bar{N}$ |
|---|---|---|---|---|---|---|---|---|---|---|---|---|---|---|---|
| | | | | \multicolumn{10}{c|}{% dose declared as MTD & average patient allocation} | | |
| | E | | Sel | 0 | 4.0 | 39.3 | **26.8** | 17.5 | 8.8 | 1.5 | 0.5 | 0.4 | 1.2 | | |
| | | | Pts | 0.4 | 4.2 | 7.6 | 5.2 | 1.3 | 1.9 | 0.1 | 0.2 | 0 | | 3.9 | 20.9 |
| 4 | | | pTox | 0.03 | 0.05 | 0.10 | 0.16 | **0.25** | 0.32 | 0.40 | 0.48 | 0.55 | | | |
| | Optimal | | Sel | 0 | 0.4 | 6.7 | 23.8 | **37.5** | 20.7 | 8.5 | 2.1 | 0.3 | | | |
| | A | $u_{01}=0.6$ | Sel | 0 | 0 | 9.4 | 34.6 | **34.3** | 15.2 | 4.4 | 0.4 | 0.3 | 1.4 | | |
| | | | Pts | 0.2 | 3.6 | 5.2 | 6.5 | 3.4 | 1.6 | 0.1 | 0.1 | 0 | | 3.2 | 20.7 |
| | | $u_{01}=0.2$ | Sel | 0 | 0 | 10.3 | 34.9 | **36.8** | 16.6 | 0 | 0 | 0 | 1.4 | | |
| | | | Pts | 0.2 | 3.6 | 5.3 | 6.7 | 3.6 | 1.5 | 0 | 0 | 0 | | 3.1 | 20.9 |
| | B | $u_{01}=0.6$ | Sel | 0 | 0 | 9.5 | 34.0 | **35.5** | 16.4 | 3.0 | 0.1 | 0.1 | 1.4 | | |
| | | | Pts | 0.2 | 3.6 | 5.2 | 6.6 | 2.6 | 2.1 | 0.4 | 0 | 0.1 | | 3.3 | 20.8 |
| | | $u_{01}=0.2$ | Sel | 0 | 0 | 10.1 | 35.7 | **37.3** | 13.8 | 1.1 | 0.5 | 0.1 | 1.4 | | |
| | | | Pts | 0.2 | 3.6 | 5.2 | 6.5 | 2.8 | 2.0 | 0.4 | 0 | 0.1 | | 3.3 | 20.8 |
| | C | | Sel | 1.2 | 0 | 0.3 | 13.3 | **34.6** | 27.9 | 20.0 | 1.5 | 0.9 | 0.3 | | |
| | | | Pts | 0 | 3.6 | 6.0 | 5.8 | 2.6 | 2.3 | 0.5 | 0 | 0.1 | | 3.4 | 20.9 |
| | D | | Sel | 0 | 0 | 0 | 2.0 | **43.6** | 53.7 | 0.7 | 0 | 0 | 0 | | |
| | | | Pts | 0 | 3.0 | 3.6 | 9.7 | 4.6 | 0.1 | 0 | 0 | 0 | | 3.2 | 21.0 |
| | E | | Sel | 0 | 2.0 | 17.1 | 21.6 | **25.0** | 20.1 | 9.9 | 1.5 | 1.5 | 1.3 | | |
| | | | Pts | 0.4 | 3.8 | 5.5 | 5.2 | 1.7 | 3.1 | 0.3 | 0.6 | 0.1 | | 3.4 | 20.7 |
| 5 | | | pTox | 0.001 | 0.005 | 0.03 | 0.10 | 0.16 | **0.25** | 0.38 | 0.50 | 0.60 | | | |
| | Optimal | | Sel | 0 | 0 | 0.1 | 8.4 | 24.6 | **44.2** | 20.4 | 2.2 | 0.1 | | | |
| | A | $u_{01}=0.6$ | Sel | 0 | 0 | 0.7 | 12.4 | 38.5 | **38.5** | 7.8 | 1.8 | 0.3 | 0 | | |
| | | | Pts | 0 | 3.1 | 3.5 | 5.1 | 4.9 | 3.5 | 0.4 | 0.4 | 0.1 | | 2.7 | 21.0 |
| | | $u_{01}=0.2$ | Sel | 0 | 0 | 0.8 | 11.8 | 43.2 | **44.2** | 0 | 0 | 0 | 0 | | |
| | | | Pts | 0 | 3.1 | 3.5 | 5.3 | 5.4 | 3.7 | 0 | 0 | 0 | | 2.5 | 21.0 |
| | B | $u_{01}=0.6$ | Sel | 0 | 0 | 0.7 | 13.6 | 38.8 | **40.3** | 6.1 | 0.1 | 0.4 | 0 | | |
| | | | Pts | 0 | 3.1 | 3.5 | 5.3 | 3.6 | 4.3 | 1.1 | 0 | 0.2 | | 2.9 | 21.0 |
| | | $u_{01}=0.2$ | Sel | 0 | 0 | 0.7 | 14.8 | **43.5** | 37.4 | 2.7 | 0.5 | 0.4 | 0 | | |
| | | | Pts | 0 | 3.1 | 3.5 | 5.2 | 3.7 | 4.2 | 1.0 | 0 | 0.2 | | 2.9 | 20.9 |
| | C | | Sel | 0 | 0 | 1.2 | 16.3 | 32.4 | **41.1** | 6.7 | 1.2 | 1.1 | 0 | | |
| | | | Pts | 0 | 3.1 | 3.7 | 5.3 | 3.2 | 4.3 | 1.1 | 0.1 | 0.2 | | 2.9 | 21.0 |
| | D | | Sel | 0 | 0 | 0 | 12.0 | **81.1** | 6.9 | 0 | 0 | 0 | 0 | | |
| | | | Pts | 0 | 3.0 | 3.0 | 6.8 | 7.8 | 0.4 | 0 | 0 | 0 | | 2.1 | 21.0 |
| | E | | Sel | 0 | 0 | 2.5 | 11.0 | 28.7 | **35.3** | 17.9 | 1.7 | 2.9 | 0 | | |
| | | | Pts | 0.1 | 3.1 | 3.7 | 5.0 | 2.0 | 5.2 | 0.7 | 1.1 | 0.1 | | 3.1 | 21.0 |
| 6 | | | pTox | 0.01 | 0.02 | 0.05 | 0.08 | 0.11 | 0.14 | **0.25** | 0.37 | 0.47 | | | |
| | Optimal | | Sel | 0 | 0 | 0.6 | 3.2 | 7.6 | 15.1 | **49.5** | 20.6 | 3.4 | | | |
| | A | $u_{01}=0.6$ | Sel | 0 | 0 | 1.4 | 7.5 | 25.1 | 37.5 | **20.4** | 6.4 | 1.3 | 0.4 | | |
| | | | Pts | 0 | 3.2 | 3.9 | 4.8 | 3.9 | 3.6 | 0.6 | 0.7 | 0.2 | | 2.1 | 20.9 |
| | | $u_{01}=0.2$ | Sel | 0 | 0 | 1.4 | 8.2 | 26.5 | 63.5 | **0** | 0 | 0 | 0.4 | | |
| | | | Pts | 0 | 3.2 | 3.9 | 4.8 | 4.5 | 4.5 | 0 | 0 | 0 | | 1.8 | 20.9 |
| | B | $u_{01}=0.6$ | Sel | 0 | 0 | 1.4 | 7.6 | 22.2 | 44.1 | **21.9** | 0.1 | 2.3 | 0.4 | | |
| | | | Pts | 0 | 3.2 | 3.9 | 4.8 | 2.2 | 4.4 | 1.7 | 0.2 | 0.5 | | 2.3 | 20.9 |
| | | $u_{01}=0.2$ | Sel | 0 | 0 | 1.4 | 8.5 | 27.0 | 46.4 | **10.6** | 3.5 | 2.2 | 0.4 | | |
| | | | Pts | 0 | 3.2 | 3.9 | 4.8 | 2.5 | 4.1 | 1.8 | 0.2 | 0.5 | | 2.3 | 21.0 |
| | C | | Sel | 0 | 0 | 1.1 | 7.4 | 18.0 | 45.7 | **17.1** | 5.1 | 5.6 | 0 | | |
| | | | Pts | 0 | 3.2 | 4.1 | 4.4 | 2.4 | 4.5 | 1.6 | 0.2 | 0.6 | | 2.3 | 21.0 |
| | D | | Sel | 0 | 0 | 0 | 9.8 | 73.6 | 16.6 | **0** | 0 | 0 | 0 | | |

Table S3 – *Continued.*

| Sc. | Design | | | $d_{i\star 1}$ 2 | $d_{i\star 2}$ 4 | $d_{i\star 3}$ 8 | $d_{i\star 4}$ 16 | $d_{i\star 5}$ 22 | $d_{i\star 6}$ 28 | $d_{i\star 7}$ 40 | $d_{i\star 8}$ 54 | $d_{i\star 9}$ 70 | None | DLT | $\bar{N}$ |
|---|---|---|---|---|---|---|---|---|---|---|---|---|---|---|---|
| | | | Pts | 0 | 3.0 | 3.1 | 6.4 | 7.8 | 0.7 | 0 | 0 | 0 | | 1.7 | 21.0 |
| | E | | Sel | 0 | 0.1 | 3.1 | 4.5 | 11.1 | 31.8 | **27.9** | 8.0 | 13.2 | 0.3 | | |
| | | | Pts | 0.2 | 3.2 | 4.0 | 4.0 | 1.1 | 4.7 | 1.2 | 2.0 | 0.5 | | 2.6 | 20.9 |
| 7 | | | pTox | 0.35 | 0.42 | 0.60 | 0.75 | 0.82 | 0.88 | 0.91 | 0.94 | 0.97 | | | |
| | Optimal | | Sel | 93.9 | 5.7 | 0.4 | 0 | 0 | 0 | 0 | 0 | 0 | | | |
| | A | $u_{01}=0.6$ | Sel | 3.4 | 6.4 | 1.4 | 0 | 0 | 0 | 0 | 0 | 0 | **88.8** | | |
| | | | Pts | 1.5 | 5.1 | 1.5 | 0.1 | 0 | 0 | 0 | 0 | 0 | | 3.6 | 8.2 |
| | | $u_{01}=0.2$ | Sel | 3.4 | 6.4 | 1.4 | 0 | 0 | 0 | 0 | 0 | 0 | **88.8** | | |
| | | | Pts | 1.5 | 5.1 | 1.5 | 0.1 | 0 | 0 | 0 | 0 | 0 | | 3.6 | 8.2 |
| | B | $u_{01}=0.6$ | Sel | 3.4 | 6.4 | 1.4 | 0 | 0 | 0 | 0 | 0 | 0 | **88.8** | | |
| | | | Pts | 1.5 | 5.1 | 1.5 | 0.1 | 0 | 0 | 0 | 0 | 0 | | 3.6 | 8.2 |
| | | $u_{01}=0.2$ | Sel | 3.4 | 6.4 | 1.4 | 0 | 0 | 0 | 0 | 0 | 0 | **88.8** | | |
| | | | Pts | 1.5 | 5.1 | 1.5 | 0.1 | 0 | 0 | 0 | 0 | 0 | | 3.6 | 8.2 |
| | C | | Sel | 4.1 | 6.9 | 0.7 | 0 | 0 | 0 | 0 | 0 | 0 | **88.3** | | |
| | | | Pts | 1.0 | 6.1 | 1.5 | 0.1 | 0 | 0 | 0 | 0 | 0 | | 3.9 | 8.7 |
| | D | | Sel | 0.1 | 59.6 | 40.3 | 0 | 0 | 0 | 0 | 0 | 0 | **0** | | |
| | | | Pts | 0 | 5.8 | 14.8 | 0.4 | 0 | 0 | 0 | 0 | 0 | | 11.7 | 21.0 |
| | E | | Sel | 6.0 | 5.1 | 0 | 0 | 0 | 0 | 0 | 0 | 0 | **88.9** | | |
| | | | Pts | 2.1 | 4.6 | 1.0 | 0.1 | 0 | 0 | 0 | 0 | 0 | | 3.3 | 7.8 |
| 8 | | | pTox | 0.001 | 0.005 | 0.01 | 0.02 | 0.04 | 0.05 | 0.10 | 0.16 | **0.25** | | | |
| | Optimal | | Sel | 0.3 | 0 | 0 | 0 | 0.6 | 0.6 | 9.2 | 29.4 | **59.9** | | | |
| | A | $u_{01}=0.6$ | Sel | 0 | 0 | 0 | 0.3 | 3.3 | 22.6 | 27.9 | 25.6 | **20.3** | 0 | | |
| | | | Pts | 0 | 3.1 | 3.1 | 3.3 | 3.5 | 4.1 | 1.1 | 1.5 | 1.3 | | 1.1 | 21.0 |
| | | $u_{01}=0.2$ | Sel | 0 | 0 | 0 | 0.3 | 4.3 | 95.4 | 0 | 0 | **0** | 0 | | |
| | | | Pts | 0 | 3.1 | 3.1 | 3.4 | 3.9 | 7.5 | 0 | 0 | 0 | | 0.6 | 21.0 |
| | B | $u_{01}=0.6$ | Sel | 0 | 0 | 0 | 0.3 | 3.0 | 24.4 | 38.0 | 0.5 | **33.8** | 0 | | |
| | | | Pts | 0 | 3.1 | 3.1 | 3.4 | 0.9 | 4.6 | 2.8 | 0.7 | 2.4 | | 1.4 | 21.0 |
| | | $u_{01}=0.2$ | Sel | 0 | 0 | 0 | 0.3 | 4.3 | 30.2 | 20.7 | 11.0 | **33.5** | 0 | | |
| | | | Pts | 0 | 3.1 | 3.1 | 3.3 | 1.1 | 4.3 | 3.0 | 0.7 | 2.4 | | 1.4 | 21.0 |
| | C | | Sel | 0 | 0 | 0 | 0.1 | 2.1 | 23.5 | 26.2 | 13.3 | **34.8** | 0 | | |
| | | | Pts | 0 | 3.0 | 3.1 | 3.3 | 0.9 | 4.5 | 3.1 | 0.7 | 2.4 | | 1.4 | 21.0 |
| | D | | Sel | 0 | 0 | 0 | 0.3 | 46.2 | 53.5 | 0 | 0 | **0** | 0 | | |
| | | | Pts | 0 | 3.0 | 3.0 | 3.7 | 9.4 | 1.9 | 0 | 0 | 0 | | 0.6 | 21.0 |
| | E | | Sel | 0 | 0 | 0 | 0.3 | 0.6 | 6.8 | 21.1 | 13.2 | **58.0** | 0 | | |
| | | | Pts | 0.1 | 3.0 | 3.1 | 3.2 | 0.3 | 4.1 | 1.5 | 3.1 | 2.6 | | 1.6 | 21.0 |

**Sc.**: Scenarios; **pTox**: true probability of toxicity in humans; **Sel**: proportion of times of declaring a dose as MTD; **Pts**: average number of patients allocated to a dose.



\end{document}